\documentclass[letterpaper,11pt,fleqn]{article}
\pdfoutput=1
\usepackage{jheppub}

\usepackage{graphicx}
\usepackage[figuresright]{rotating}
\usepackage{bm,amsmath,amssymb}
\usepackage[mathscr]{eucal}
\usepackage{color}

\long\def\comment#1{ }
\newcommand{\eqn}[1]{Eq.~\eqref{#1}}
\newcommand{\beq}{\begin{equation}}
\newcommand{\eeq}{\end{equation}}
\newcommand{\nn}{\nonumber\\}

\newcommand{\rmd}{{\rm d}}
\newcommand{\rme}{{\rm e}}

\newcommand{\del}{\partial}

\newcommand{\order}[1]{\mcal{O}{\left(#1\right)}}
\newcommand{\mcal}{\mathcal}

\newcommand{\vep}{\varepsilon}

\newcommand{\abar}{\bar{\alpha}_s}

\newcommand{\tbr}{t_{\rm br}}
\newcommand{\tf}{t_{\rm f}}
\newcommand{\obr}{\omega_{\rm br}}

\title{\Large Event-by-event fluctuations in the medium-induced jet evolution} 

\author{Miguel A.~Escobedo}

\author{and Edmond Iancu}

\affiliation{Institut de physique th\'{e}orique, Universit\'{e} Paris Saclay, CNRS, CEA, F-91191 Gif-sur-Yvette, France}

\emailAdd{Miguel-Angel.Escobedo-Espinosa@cea.fr}
\emailAdd{Edmond.Iancu@cea.fr}

\abstract{We develop the event-by-event picture of the gluon distribution produced 
via medium-induced gluon branching
by an energetic jet which propagates through a dense QCD medium.
A typical event is characterized by the production of a large number
of soft gluons which propagate at large angles with respect to the jet axis and
which collectively carry a substantial amount of energy. 
By explicitly computing 2-gluon correlations, we demonstrate the existence of large 
event-by-event fluctuations, which reflect the stochastic nature of the branching process.
For the two quantities that we have investigated ---
the energy loss at large angles and  the soft gluon multiplicity  ---, 
the dispersion is parametrically as large as the respective expectation value. 
We identify interesting scaling laws, which suggest
that the multiplicity distribution should exhibit KNO
(Koba-Nielsen-Olesen) scaling. A similar scaling is known to hold for a jet branching
in the vacuum, but the medium-induced distribution is found to be considerably broader.
We predict that event-by-event measurements of the di-jet asymmetry in Pb+Pb
collisions at the LHC should
observe large fluctuations in the number of soft hadrons propagating at large angles
and also in the total energy carried by these hadrons.

}


\begin{document}
\maketitle

\section{Introduction}
\label{sec:intro}

A remarkable phenomenon discovered and, by now, abundantly studied in
Pb+Pb collisions at the LHC is the {\em di-jet asymmetry} --- a strong energy imbalance
between two jets which propagate nearly back-to-back in the plane transverse 
to the collision axis  \cite{Aad:2010bu,Chatrchyan:2011sx,Chatrchyan:2012nia,Aad:2012vca,Chatrchyan:2013kwa,Chatrchyan:2014ava,Aad:2014wha,Khachatryan:2015lha,Khachatryan:2016erx}. 
Imbalanced di-jets are
also observed in proton-proton collisions, where they are mostly associated with 3-jet events,
but the respective events in (central) Pb+Pb collisions look quite different. First, 
the fraction of events showing a strong energy asymmetry is considerably higher. 
Second, the pattern of the asymmetry, i.e. the distribution of the `missing' energy
in the underlying event, is very different: in heavy ion collisions, the energy difference 
between the leading and the subleading jets is not carried by a 3rd jet anymore, 
but rather by many soft hadrons, with energies $p_T\le 2$\,GeV, which propagate 
in the hemisphere of the subleading jet at large angles w.r.t. the latter.
Accordingly, the `missing' energy is only slowly recovered when 
increasing the angular opening $\Delta R$ of the subleading jet; for instance, for the
relatively large value $\Delta R=0.8$, there is still a substantial energy imbalance,
$\Delta E\simeq 20\div 30$\,GeV  \cite{Chatrchyan:2011sx,Khachatryan:2015lha}.

It seems natural to associate the di-jet asymmetry in Pb+Pb collisions with
the interactions between the partons in the jets and the surrounding 
QCD medium --- the quark-gluon plasma --- which is expected to be
created in the intermediate stages of a ultrarelativistic nucleus-nucleus collision.
Such interactions (e.g. elastic collisions) can easily deviate the soft components of a jet 
towards large angles  and an asymmetry is generated whenever one of the two 
jets crosses the medium along a longer distance than the other
\cite{CasalderreySolana:2010eh}. However, the main question remains: how is that
possible that a substantial fraction of the energy of a jet is carried by its {\em soft}
constituents ? As well known, bremsstrahlung in QCD cannot lead to an efficient redistribution
of the energy towards soft particles: while this mechanism favors the emission of soft gluons,
the total energy carried by these gluons remains negligible. This is so since each
radiated gluon carries only a tiny fraction $x\ll 1$ of the energy of its parent parton.

However, as pointed out in Ref.~\cite{Blaizot:2013hx}, the evolution of a jet in a dense
medium is dramatically different from that that would be generated by bremsstrahlung in 
the vacuum: at sufficiently low energies, the BDMPSZ rate for medium-induced gluon 
branching favors {\em multiple branching} which is {\em quasi-democratic}. The
BDMPSZ rate (from Baier, Dokshitzer, Mueller, Peign\'e, Schiff, and Zakharov)  
\cite{Baier:1996kr,Baier:1996sk,Zakharov:1996fv,Zakharov:1997uu,Baier:1998kq} 
refers to gluon emissions which are triggered by multiple soft scattering between
the gluon which is about to be emitted and the constituents of the medium (see also 
\cite{Wiedemann:2000za,Wiedemann:2000tf,Arnold:2001ba,Arnold:2001ms,Arnold:2002ja}
for related developments). Unlike the rate for bremsstrahlung, the BDMPSZ rate depends 
also upon the energy of the parent parton (and not only upon the energy fraction $x$ of the 
emitted gluon). Namely, this rate becomes large when the emitter is sufficiently soft
(irrespective of the value of $x$) and thus favors the development of democratic
cascades via the successive branchings of soft gluons. The democratic branching
is a very efficient mechanism for transmitting the energy from the leading particle
(the energetic parton which has initiated the jet) to a myriad of soft gluons, which are
eventually deviated to large angles by rescattering in the medium.


As demonstrated in Refs.~\cite{Blaizot:2012fh,Blaizot:2013vha,Apolinario:2014csa},
the medium-induced evolution via multiple branching is a classical stochastic process,
which is {\em Markovian}\footnote{This property was implicitly assumed in earlier
works aiming at constructing a kinetic theory for in-medium jet evolution \cite{Baier:2000sb,Baier:2001yt,Arnold:2002zm,Jeon:2003gi,Schenke:2009gb}. The study of color decoherence in a 
dense partonic medium has been pioneered in Refs.~\cite{MehtarTani:2010ma,MehtarTani:2011tz,CasalderreySolana:2011rz} and the precise conditions for the factorization
of successive emissions have been clarified in Refs.~\cite{Blaizot:2012fh,Apolinario:2014csa}. 
See also Refs.~\cite{CasalderreySolana:2012ef,Kurkela:2014tla,Casalderrey-Solana:2015bww} 
for related studies and the recent review paper \cite{Blaizot:2015lma}.}: 
after each splitting, the two daughter gluons evolve independently
from each other because their color coherence is rapidly washed out
by rescattering in the medium. So far, only the {\em mean field}
aspects of this stochastic evolution, as encoded in the average gluon spectrum 
$D(x)=x({\rmd N}/{\rmd x})$ and the associated transport equation, have been explicitly 
studied. For the branching dynamics alone, one has obtained exact solutions for
$D(x)$ \cite{Blaizot:2013hx,Fister:2014zxa,Blaizot:2015jea}, which demonstrate the 
physics of democratic branchings and the associated phenomenon of {\em wave turbulence}
(the energy flux down the democratic cascades is uniform in $x$).
Using more general equations which include the elastic collisions with the medium constituents, 
one has also studied 
the interplay between branchings and elastic collisions, and its consequences for the angular
structure of the cascade
 \cite{Blaizot:2013vha,Iancu:2014aza,Kurkela:2014tla,Blaizot:2014ula,Blaizot:2014rla} and for the 
thermalization of the soft branching products \cite{Iancu:2015uja}.

In this paper, we shall for the first time investigate the {\em event-by-event fluctuations}
introduced by the branching process in the medium-induced jet evolution. Specifically,
we shall compute the dispersion in the energy lost by the jet at large angles and also
in the number of gluons $N(x_0)$ having an energy fraction $x$ larger than some infrared
cutoff $x_0\ll 1$. Here, $x\equiv\omega/E$, with $E$ the energy of the leading particle
and $\omega\le E$ the energy of a parton from the jet. We shall consider a pure branching
process, without elastic collisions, so our results strictly apply only for sufficiently high energies
$\omega \gg T$, with $T$ the typical energy of the medium constituents. (The medium is 
assumed to be a weakly-coupled quark-gluon plasma in thermal equilibrium at temperature $T$;
for the present purposes, it is fully characterized by a transport coefficient for transverse 
momentum diffusion, known as the `jet quenching parameter' $\hat q$.)
This is not a serious limitation so long
as we are interested only in the overall energy transmitted by the jet to the medium, and not
also in its detailed distribution in space and time \cite{Iancu:2015uja}. 
For this branching process, we shall obtain an exact result, \eqn{D2exact}, for the gluon
pair density $D^{(2)}(x,x')\equiv xx'(\rmd N/\rmd x\rmd x')$. Using this result,
we shall compute the aforementioned dispersions. 

Our main conclusion is that
{\em fluctuations are large}: for both the energy loss at large angles and the gluon 
multiplicity $N(x_0)$, the dispersion is parametrically as large as the respective average
quantity. Such large fluctuations should be easy to observe in event-by-event studies
of the di-jet asymmetry at the LHC. In particular, we predict that the multiplicity fluctuations 
should exhibit Koba-Nielsen-Olesen (KNO) scaling \cite{Koba:1972ng}.
A similar scaling is known to hold for a jet branching in the vacuum  \cite{Dokshitzer:1991wu}, 
but the medium-induced gluon
distribution is found to be considerably wider (see Sect.~\ref{sec:number} for details).

The physical picture emerging from our analysis can be summarized as follows.
In a typical event, the leading particle evolves by emitting a number of order one of primary 
gluons with energy $\omega\sim \obr(L)$ together with a large number
$\sim [\obr(L)/\omega]^{1/2}$  of considerably softer
gluons, with $\omega\ll\obr(L)$. Here, $\obr(L)\equiv \abar\hat q L^2$, with
$\abar=\alpha_s N_c/\pi$ and $L$ the distance travelled by the jet through the medium,
is the characteristic energy scale for medium-induced multiple branching.  (We implicitly
assume here that $\obr(L)\ll E$, as this is the typical situation for high energy jets at the LHC;
see the discussion at the end of Sect.~\ref{sec:phys}.) The primary gluons 
with $\omega\lesssim \obr(L)$ develop mini-jets via successive democratic branchings and
thus transmit their whole energy to softer quanta with $\omega\sim T$ that are expected
to thermalize via collisions in the medium \cite{Iancu:2015uja}. Harder emissions with
$\omega\gg\obr(L)$ occur only rarely, with a small probability $\sim [\obr(L)/\omega]^{1/2}$,
and moreover they do not contribute to the energy lost by
the jet as a whole, since hard gluons propagate at small angles.
The energy loss at large angles is rather controlled by the {\em hardest typical emissions},
those with energies $\omega\sim \obr(L)$. As aforementioned, the number of
such emissions is of order one and they occur independently from each other;
accordingly, both the average energy loss by the jet and its dispersion
are of order $\obr(L)$ (see \eqn{dispersion}).

Unlike the primary gluons (the direct emissions by the leading particle), which are
quasi independent from each other, the gluons from the subsequent generations, as produced
via democratic branchings, can be mutually correlated due to the fact that they have common
ancestors. This correlation is particularly important for the sufficiently soft gluons,
with  $\omega\ll \obr(L)$, which belong to a same mini-jet. It is responsible for the strong
fluctuations in the gluon multiplicity at small $x$, leading to the KNO scaling
alluded to above (see \eqn{NNcum} and the subsequent discussion).

Another interesting conclusion of our analysis is that the gluon distribution at small $x$
is not affected by multiple branching: both the gluon spectrum $D(x)$
and the gluon pair density $D^{(2)}(x,x')$ are formally the same, when $x,\,x'\ll 1$,
as the respective predictions of the lowest-order perturbative expansion in the number of branchings.
For the gluon spectrum, this property was already known \cite{Baier:2000sb,Blaizot:2013hx}: 
this is the statement that the low-$x$ limit of the BDMPSZ spectrum, namely the power-law
spectrum $D(x)\propto 1/\sqrt{x}$, 
is a fixed point of the branching dynamics. The present analysis shows that
a similar property holds for the 2-gluon correlation $D^{(2)}(x,x')$ 
(see \eqn{D2small} and the related discussion).  This might look surprising for the soft
gluons with $\omega\lesssim \obr(L)$, for which multiple branching is known to be important,
but it is most surely a consequence of wave turbulence: the rate at which the soft gluons are 
produced via the splitting of harder gluons is precisely equal to the rate at which they
 disappear by decaying into even softer gluons. 
This fine balance between `gain' and `loss' ensures that the energy flux 
is independent of $x$, which is the hallmark of wave turbulence \cite{KST,Nazarenko}.

While not manifest in the gluon distribution at small $x$, the effects of multiple branchings are
visible in other aspects of this distribution, like the structure of the leading particle peak
near $x=1$, and also in {\em global} properties, like the phenomenon of wave turbulence
and its consequences for the energy loss by the jet at large angles. Such phenomena are
quite intricate and often elusive, hence the importance of disposing of {\em exact} analytic
solutions in order to unambiguously demonstrate them.
Analytic approximations and parametric estimates can be helpful for developing a 
general picture, but they cannot capture subtle aspects like the broadening of the leading
particle peak via multiple emissions, or the existence of a turbulent 
energy flux, uniform in $x$.
Numerical solutions too may alter some aspects of the dynamics, like the power-law
spectrum at small $x$, due to artifacts like infrared cutoffs. This being said, and once
a coherent physical picture has been established on the basis of exact results, the
numerical methods offer a powerful tool for extending these solutions to more realistic
situations and, especially, for systematic studies of the phenomenology. In particular,
the Monte-Carlo implementation of the in-medium jet evolution known as MARTINI 
\cite{Schenke:2009gb} should allow for
general studies of the stochastic aspects that we shall identify in this
paper and of their implications for the phenomenology at RHIC and the LHC.

This paper is organized as follows. In Sect.~\ref{sec:fluct} we succinctly describe the
physical picture of the medium-induced jet evolution and its mathematical formulation
as a Markovian process. In particular, we present the transport equation \eqref{eqD2}
obeyed by the gluon pair density $D^{(2)}(x,x')$. More details on the formalism are
deferred to App.~\ref{app:Markov}. In Sect.~\ref{sec:var} we discuss the energy loss
at large angles, operationally defined as the total energy transmitted, via
successive branchings, to the very soft gluons with $x\to 0$.
Sect.~\ref{sec:spec} is devoted to the mean field picture, 
that is, the gluon spectrum $D(x)$ and the average energy loss.
Most of the results presented there were already known, but our 
physical discussion is more furnished, in line with our general purposes. 
In Sect.~\ref{sec:pair}, we present our main
new results, which are both exact (within our theoretical framework): 
\eqn{D2exact} for the gluon pair density $D^{(2)}(x,x')$ and
\eqn{VAR} for the variance in the energy loss at large angles. The physical 
interpretation of these results is discussed at length, in Sect.~\ref{sec:pair} 
and the dedicated section \ref{sec:dis}. Details on the calculations are presented in Appendices \ref{app:D2}
and  \ref{app:X2}. In Sect.~\ref{sec:number}, we discuss the gluon number distribution,
for gluons with energy fraction $x\ge x_0$. In Sect.~\ref{sec:avn} we compute
the average multiplicity, while in Sect.~\ref{sec:NN} we present and discuss our results 
for the second factorial moment $\langle N(N-1)\rangle$ and for the variance. 
The respective calculations are quite tedious (the details are deferred to App.~\ref{app:N2}),
but in Sect.~\ref{sec:NN} we anticipate the final results via physical considerations, which also 
shed light on their remarkable scaling properties.
Sect.~\ref{sec:conclusion} presents a brief summary of our results together with possible implications
for the phenomenology.


\section{Medium-induced jet evolution: the general picture}
\label{sec:fluct}

As explained in the Introduction, we consider the jet generated via medium-induced
gluon branching by a leading particle (LP)
with initial energy $E$ which propagates along a distance $L$ through a weakly coupled
quark-gluon plasma with temperature $T\ll E$. Some typical values, as
inspired by the phenomenology at the LHC, are $E=100$\,GeV and $T=0.5$\,GeV.

We would like to study the event-by-event distribution of the energy lost by the jet
at large angles with respect to the jet axis, say at polar angles 
$\theta>\theta_0$ with $\theta_0\sim \order{1}$. 
To that aim, one needs the distribution of the radiated gluons
in energy ($\omega$) and polar angle ($\theta$), or, equivalently, in $\omega$  and
transverse momentum $k_\perp$ (recall the trigonometric relation $\sin\theta=k_\perp/\omega$).
For simplicity though, and in order to allow for analytic studies,
we shall explicitly consider only the distribution in $\omega$, as obtained after integrating
out $k_\perp$. Then the angular distribution will be approximately reconstructed  by
using the fact that, for medium-induced emissions,
the $k_\perp$-distribution is strongly peaked at\footnote{This estimate
for $k_\perp^2$ ceases to be valid for the very soft gluons with energies $\omega\ll\obr(L)$,
whose lifetimes $\tbr(\omega)$ are much shorter than $L$ (see \eqn{tbr}).
For such gluons, $k_\perp^2\simeq \hat q\tbr(\omega)$ is smaller than $Q_s^2$
and also $\omega$-dependent \cite{Iancu:2014aza,Kurkela:2014tla,Blaizot:2014ula}. 
However, this brings no serious complication for our present
arguments, since the energy loss  in \eqn{Ex0} is quasi-independent  of $x_0$, hence
of $\omega_0$, whenever $\omega_0\ll \obr(L)$; see Fig.~\ref{fig:X} and the associated discussion.}
 $k_\perp^2= Q_s^2\equiv\hat q L$.
Indeed, this is the typical transverse momentum acquired via medium rescattering
by a gluon propagating through the medium along a distance $L$.
For the phenomenologically 
relevant values $\hat q= 1$\,GeV$^2/$fm  and $L=4$\,fm, one finds $Q_s\simeq 2$\,GeV,
meaning that gluons with energies $\omega\le 2$\,GeV will propagate
at angles  $\theta\gtrsim 1$.  

The above considerations motivate the
following estimate for the energy which propagates at
angles $\theta>\theta_0$ at time $t$, with $0\le t\le L$ :
\beq\label{Ex0}
\mcal{E}(t,x_0)\,=\,E-\int_{\omega_0}^E\rmd\omega \,\omega \,\frac{\rmd N}{\rmd \omega}
\,=\,E\left(1 -\int_{x_0}^1\rmd x\,x\,\frac{\rmd N}{\rmd x}\right).
\eeq
Here, ${\rmd N}/{\rmd \omega}$ is the energy distribution of the gluons from
the jet in a given event\footnote{On an event
by event basis, ${\rmd N}/{\rmd \omega}$ is truly a discrete quantity --- the number of gluons
in each bin of $\omega$ ---, so the integrals occurring in \eqn{Ex0} should be properly viewed as
sums (see also \eqn{Dxt} below). We nevertheless use continuous notations for
convenience.},
$x\equiv \omega/E$ is the energy fraction of a gluon, $x_0\equiv \omega_0/E$, and the infrared
cutoff $\omega_0$ is the energy corresponding 
(via the simple argument above) to the limiting angle $\theta_0$~:
$\sin\theta_0=Q_s/\omega_0$.
The r.h.s. of \eqn{Ex0} is recognized as the difference between the initial energy $E$ of the 
LP and the energy carried by the components of the jet whose energies are larger
than $\omega_0$ (and hence propagate at angles smaller than $\theta_0$). 

For the jets produced in actual nucleus-nucleus collisions, this quantity $\mcal{E}(t,x_0)$ can
significantly vary from one event to another, due to various reasons. First, the medium properties
(like the value of $\hat q$) and its geometry (size and shape) can be different in
different events --- albeit in practice one can diminish the importance of such fluctuations by choosing
events in a same centrality class. Second, even for a given collision geometry,  there can be large 
fluctuations in the distance $L$ travelled by the jet through the medium, because the LP can be 
produced at any point within the interaction volume and with any initial orientation.
Finally, the evolution of the jet via successive medium-induced branchings is a
stochastic process, which is characterized by strong fluctuations, as we shall see.

In what follows, we shall only address the 
last source of fluctuations, that is, we shall assume fixed values for the physical 
parameters $\hat q$, $L$ and $E$,
and study the distribution of $\mcal{E}(x_0,t)$ generated by the stochastic branching process.
To that aim, we shall rely on the Markovian description for this process, as recently developed in 
Refs.~\cite{Blaizot:2013hx,Blaizot:2013vha,Fister:2014zxa,Blaizot:2015jea,Blaizot:2015lma}.
In this section we shall succinctly review this description, starting with the underlying physical
picture. Besides motivating the various approximations, this discussion will be also
useful for the physical interpretation of our new results.

\subsection{Democratic branchings and wave turbulence}
\label{sec:phys}

Consider a generic gluon from the medium-induced cascade, whose energy $\omega$ satisfies 
$T\ll \omega \le E$. This gluon has a probability
\beq\label{Pdeltat}
 \Delta \mcal{P}\,=\,2{\abar} \,\frac{\Delta t}{\tf (z\omega)}
\,=\, 2{\abar} \,\sqrt{\frac {\hat q}{z \omega}}\,
 \Delta t\,,\eeq
to radiate a soft gluon with energy $\omega'\ge z \omega$ 
during a time interval $\Delta t$ ($z\ll 1$ is the splitting fraction). 
In \eqn{Pdeltat}, $\abar \equiv \alpha_s N_c/\pi$,
$\hat q$ is the jet quenching parameter, and $\tf (z\omega)=\sqrt{z\omega/\hat q}$
is the quantum formation time for the emission of a gluon with energy $z\omega$.
So, \eqn{Pdeltat} can be understood as follows: $\Delta \mcal{P}$ is the product of
the probability $\abar$ for one gluon emission times the number of times that this emission
can fit within a given time interval $\Delta t$.
This probability becomes of order one, meaning that a branching is certain to occur, after a time
\beq\label{Deltat}
\Delta t(\omega, z)\,=\,
 \frac{1}{2\abar} \,\sqrt{\frac{z \omega}{\hat q}}\,.
 \eeq
This time is relatively short when $z\ll 1$, meaning that soft gluons are abundantly produced. Clearly,
the original gluon survives to such soft emissions, in the sense that it can be unambiguously 
distinguished from its radiation products, due to its higher energy.
But after the larger time interval
\beq\label{tbr}
\tbr(\omega)\,\equiv \,\frac{1}{\abar}\sqrt{\frac{\omega} {\hat q}}\,= \,\frac{1}{\abar}\,\tf(\omega)
\,,\eeq
the gluon $\omega$ is bound to undergo a `quasi-democratic' ($z\sim \order{1}$) branching 
and then it disappears --- it gets replaced by a pair of daughter gluons whose energies
are comparable to each other. In turn, these daughter gluons disappear (via democratic splittings)
even faster, since their energies are lower. This democratic branching process repeats itself until the original 
energy $\omega$ gets transmitted to a multitude of soft gluons with comparable energies.
Eventually, the dynamics changes when the energies of the soft descendants become of the
order of the medium scale $T$~: such very soft gluons can efficiently thermalize via elastic
collisions off the medium constituents and thus they transmit their whole energy to the medium
\cite{Iancu:2015uja}.

A fundamental property of a democratic cascade, known as `wave turbulence', is the fact
that the {\em energy flux} generated via democratic branchings --- the rate at which the energy 
flows across a given  bin in $\omega$ --- is independent of $\omega$,
for $\omega\ll E$ \cite{Blaizot:2013hx,Fister:2014zxa}.
Through successive branchings, the energy flows from one parton generation to the next one, 
without accumulating at any intermediate value of $\omega$. In the absence of thermalization
effects, the whole energy would eventually accumulate into a condensate at $\omega=0$ \cite{Blaizot:2013hx}. In reality, the energy is eventually
transmitted to the medium constituents, via elastic collisions leading to the
thermalization of the soft branching products with $\omega\lesssim T$ \cite{Iancu:2015uja}.
But the rate at which this happens is still controlled by the branching process 
at $\omega> T$, that is, by the turbulent flux alluded to above, which is independent of $\omega$
and hence can be formally computed at $\omega=0$.

These considerations suggest that, in order to evaluate the energy loss in
\eqn{Ex0}, one can
restrict oneself to a pure branching process (no elastic collisions) and take the limit
$x_0\to 0$. This is consistent with the fact that the energy which is physically lost towards the 
medium is the same as that which would accumulate at $x=0$ in the absence of 
thermalization. Hence, one can replace \eqn{Ex0} by
\beq\label{E0}
\mcal{E}(t)\,=\,E\left(1-X(t)\right)\,,\qquad
X(t)\equiv \int_{0}^1\rmd x\,x\,\frac{\rmd N}{\rmd x}(t)\,,
\eeq
with ${\rmd N}/{\rmd x}$ determined by the pure branching process discussed
at length in Refs.~\cite{Blaizot:2013hx,Blaizot:2013vha,Fister:2014zxa,Blaizot:2015jea,Blaizot:2015lma}.

We conclude this subsection with some remarks of relevance for the phenomenology
of high energy jets ($E\ge 100$\,GeV) at the LHC.
In discussing democratic branchings so far, we have implicitly assumed that the branching time 
\eqref{tbr} is shorter than the medium size $L$. This in turn implies an upper bound on the energy
$\omega$ of the parent gluon: $\omega \lesssim \omega_{\rm br}(L)\equiv \abar^2\hat q L^2$. 
This condition however is {\em not} satisfied by the LP in the experimental set-up at the LHC: 
its energy $E\ge 100$\,GeV is considerably larger than the typical values expected for 
$\omega_{\rm br}(L)$. (For instance, with $\abar=0.3$, $\hat q= 1$\,GeV$^2/$fm, and $L=4$\,fm, one finds
$\omega_{\rm br}\simeq 8$\,GeV.) This means that the LP cannot undergo (medium-induced)
democratic splittings and hence it emerges out of the medium, as the core of
the surviving jet. A similar argument applies to the relatively hard 
emissions\footnote{The energy of a medium-induced emission is limited by the
condition that the formation time $\tf(\omega)$ be smaller than $L$. This implies
$\omega\le \omega_c(L)\equiv \hat q L^2$  \cite{Baier:1996kr,Zakharov:1996fv}. However,
the relatively hard gluons with $\obr(L) < \omega <  \omega_c(L)$ cannot 
undergo democratic branchings and hence do not contribute to the energy loss at large angles.}
with energies $\omega \gg \omega_{\rm br}(L)$, which are rare events: such hard gluons  
survive the medium and propagate at small angles, hence they appear in the final state as 
components of the conventionally defined jet. 

On the other hand, the LP
can emit {\em abundantly}, i.e. with probability of order one, relatively soft gluons
with $\omega \lesssim \omega_{\rm br}(L)$. 
Such soft primary emissions determine
the structure of a {\em typical event}. Each of these primary gluons will generate
its own cascade, or `mini-jet', via successive democratic branchings and thus eventually
transmit its initial energy to the medium. The `hardest' among these soft gluons,
those with energies of order $ \omega_{\rm br}(L)$, will play an essential role in
what follows: they control both the energy loss at large angles and its
fluctuations, as we shall see.

\subsection{Transport equations for the gluon distribution}
\label{sec:eqs}

In this section, we shall relate the average energy loss $\langle \mcal{E}(t)\rangle$
and its variance $\langle \mcal{E}^2(t)\rangle - \langle \mcal{E}(t)\rangle^2$ to the
one- and two-point functions of the energy density $x({\rmd N}/{\rmd x})$, for
which we shall then establish evolution equations.
For the average energy loss, this relation is quite obvious:
\beq\label{avvep}
\langle \varepsilon(t)\rangle\,=\,1-\langle X(t)\rangle\,,\qquad
\langle X(t) \rangle = \int_{0}^1\rmd x\, D(x,t)\,,
\eeq
where $\vep\equiv \mcal{E}/E$ is the fraction of the total energy which accumulates
at $x=0$ and $D(x,t)$ is the average energy density, or gluon spectrum, as
defined in \eqn{correlators} below. For the variance, we have
\beq\label{sigma}
\sigma_\varepsilon^2(t)\,\equiv\,
\langle \varepsilon^2(t)\rangle-\langle \varepsilon(t)\rangle^2\,=\,
\langle X^2(t)\rangle-\langle X(t)\rangle^2\,.\eeq
As we shortly explain, the quantity $\langle X^2(t)\rangle$ can be evaluated as
\beq\label{X2}
\langle X^2(t)\rangle = \int_{0}^1\rmd x \int_{0}^1\rmd x'\, D^{(2)}(x,x',t)\,+\,
\int_{0}^1\rmd x\, x\,D(x,t)\,,
\eeq
where $D^{(2)}(x,x',t)$, with $x+x'\le 1$, is the average density of pairs of gluons multiplied by $x x'$~:
\beq\label{correlators}
D(x,t)\equiv x\,\left\langle \frac{\rmd N}{\rmd x}(t)\right\rangle\,,
\qquad
D^{(2)}(x,x',t)\equiv x x'\,\left\langle \frac{\rmd N_{\rm pair}}{\rmd x \,\rmd x'}(t)\right\rangle\,.
\eeq

To render these definitions more explicit, let us consider a more detailed description
of the statistical ensemble of events which is created by the stochastic branching process
\cite{Blaizot:2013vha}.
The central ingredient is the probability density $ \mcal{P}_n(x_1,x_2,\cdots,x_n| t)$
for having a state with $n$ gluons with energy fractions $x_i$ ($i=1,\dots,n$), at time $t$.
At $t=0$, we have just the LP, hence $ \mcal{P}_n( t=0)=\delta_{n1}\delta(x_1-1)$.
The expectation value of an arbitrary observable is computed as
\beq\label{aveO}
\langle \mcal{O}(t)\rangle\equiv
\sum_{n=1}^\infty\int \prod_{i=1}^n \rmd x_i\,  \mcal{P}_n(x_1,x_2,\cdots,x_n| t)\,
\mcal{O}_n\,,\eeq
where $\mcal{O}_n\equiv \mcal{O}(x_1,x_2,\cdots,x_n)$ denotes the value of
$\mcal{O}$ in a particular event with $n$ gluons. In particular, the quantities introduced in
\eqn{correlators} involve ${\rmd N_n}/{\rmd x}=\sum_i^n\delta(x_i-x)$, hence
\beq\label{Dxt}
D(x,t)\,=x\,\left\langle \sum_i^n\delta(x_i-x)\right\rangle\,,\qquad 
D^{(2)}(x,x',t)= x x'\,\left\langle 
\sum_{i\ne j}^n\delta(x_i-x) \delta(x_j-x')\right\rangle\,.\eeq
Note the distinction $i\ne j$ in the definition
of $D^{(2)}(x,x',t)$, which shows that this quantity is indeed the
density of {\em pairs} of gluons, each pair being counted twice (since
the pairs $(x,x')$ and $(x',x)$ are separately counted).
Using \eqn{Dxt}, it is easy to check the previous formula
\eqref{X2}.

At this point, it is important to observe that this probabilistic description requires 
an infrared cutoff (say, a lower
limit on $x$), playing the role of an energy resolution scale, below which
gluons cannot be resolved anymore. Indeed, the branching dynamics produces an
infinite number of arbitrarily soft gluons and the `state with exactly $n$ gluons' is not well defined
without such a cutoff.  Any explicit construction of such a state, say via Monte-Carlo 
simulations, must involve an infrared cutoff on $x$, to be viewed as a part of the `state' definition.
On the other hand, physical quantities which are sufficiently inclusive, such as
$D(x,t)$ and $D^{(2)}(x,x',t)$, are insensitive to the unobserved, soft, gluons, hence they
must be independent of this cutoff. And indeed, one can show that the cutoff dependence
cancels out in the equations obeyed by these quantities. So long as one is solely 
interested in such quantities, one can formally proceed without any infrared
cutoff (see the discussion in App.~\ref{app:Markov}).

The probability densities $ \mcal{P}_n(x_1,x_2,\cdots,x_n| t)$ obey `master equations', that is, coupled
rate equations whose general structure is rather standard, since common to all the Markovian 
branching processes. (These equations are presented in App.~\ref{app:Markov} for convenience;
see also \cite{Blaizot:2013vha}.)
What is specific for the problem at hand, is the expression of the splitting
rate, here given by the BDMPSZ mechanism \cite{Baier:1996kr,Zakharov:1996fv}.
Namely, the differential probability per unit time and per unit $z$ for the splitting
of the parent gluon $\omega=xE$ into the pair of daughter gluons $z\omega$ 
and $(1-z)\omega$ can be conveniently written as
 \beq\label{Pdef}
 \frac{\rmd^2 \mcal{I}_{\rm br}}{\rmd z\,\rmd t}
 \,=\,\frac{\mcal{K}(z)}{2\,\tbr(\omega)} \,=\,\frac{1}{2\,\tbr(E)}\,\frac{\mcal{K}(z)}{\sqrt{x}}\,,
 \qquad {\cal K}(z)\equiv\frac{[1-z(1-z)]^{\frac{5}{2}}}{[z(1-z)]^{\frac{3}{2}}}\,,
  \eeq
with $\tbr(\omega)$  the branching time from  \eqn{tbr}. Using the master equations
within \eqn{aveO}, it is straightforward to derive transport equations for the observables
(see App.~\ref{app:Markov} for details).
The corresponding equation for the gluon spectrum $D(x,t)$ reads\footnote{This is a simplified
version of equations previous used in Refs. \cite{Baier:2000sb,Jeon:2003gi,Schenke:2009gb},
which in turn represent the branching
part of the more general kinetic equation constructed in \cite{Arnold:2002zm} 
for the case of a weakly-coupled quark-gluon plasma.}
 \beq\label{eqD}
\frac{\partial}{\partial\tau} D(x,\tau)=\int \rmd z \,
{\cal K}(z)\left[\sqrt{\frac{z}{x}}D\left(\frac{x}{z},\tau\right)-\frac{z}{\sqrt{x}}D(x,\tau)\right],
\eeq
where we introduced the dimensionless time variable $\tau\equiv t/\tbr(E)$, that is, the
physical time in units of the branching time $\tbr(E)$ of the LP. 
The first term in the r.h.s. describes the gain in the number 
of gluons at  $x$ due to emissions from gluons with $x'=x/z > x$, 
whereas the second term describes the loss via the decay into softer gluons.
Taken separately, the gain term and the loss term
have endpoint singularities at $z=1$ (corresponding to the
emission of very soft gluons), but these singularities mutually
cancel and the overall equation is well defined without any infrared
cutoff, as anticipated.

The corresponding equation for the two-point function will be derived in App.~\ref{app:Markov},
and reads
\begin{align}\label{eqD2}
   \frac{\del }{\del\tau}D^{(2)}(x,x',\tau)
  &=\int \rmd z \,
  {\cal K}(z)\left[\sqrt{\frac{z}{x}}D^{(2)}\Big(\frac{x}{z}, x',\tau\Big)-\frac{z}{\sqrt{x}}D^{(2)}\big({x},x',\tau\big)\right]
  \, + \,\Big( x \,\leftrightarrow\, x'\Big)\nonumber\\*[0.2cm]
  &\qquad +\,\frac{x x'}{(x+x')^2}\,
   {\cal K}\Big(\frac{x}{x+x'}\Big)\,\frac{1}{\sqrt{x+x'}}\,D(x+x',\tau)\,.
  \end{align}
The two terms in the r.h.s. of the first line are easy to understand by comparison with \eqn{eqD}:
they describe the separate evolutions in the numbers of gluons
with energy fractions $x$ and $x'$, respectively.
The  `source' term in the second line describes the simultaneous creation of a pair of gluons with energy
fractions $x$ and $x'$ via the branching of a parent gluon with energy fraction $x+x'$
(with $x+x' \le 1$ of course).
This terms introduces correlations in the evolution of the pair of gluons and is responsible for 
the fluctuations to be analyzed in the next sections.

\begin{figure}[t]
\begin{minipage}[b]{0.50\textwidth}
\begin{center}
\vspace*{-1.cm}
\includegraphics[width=0.95\textwidth]{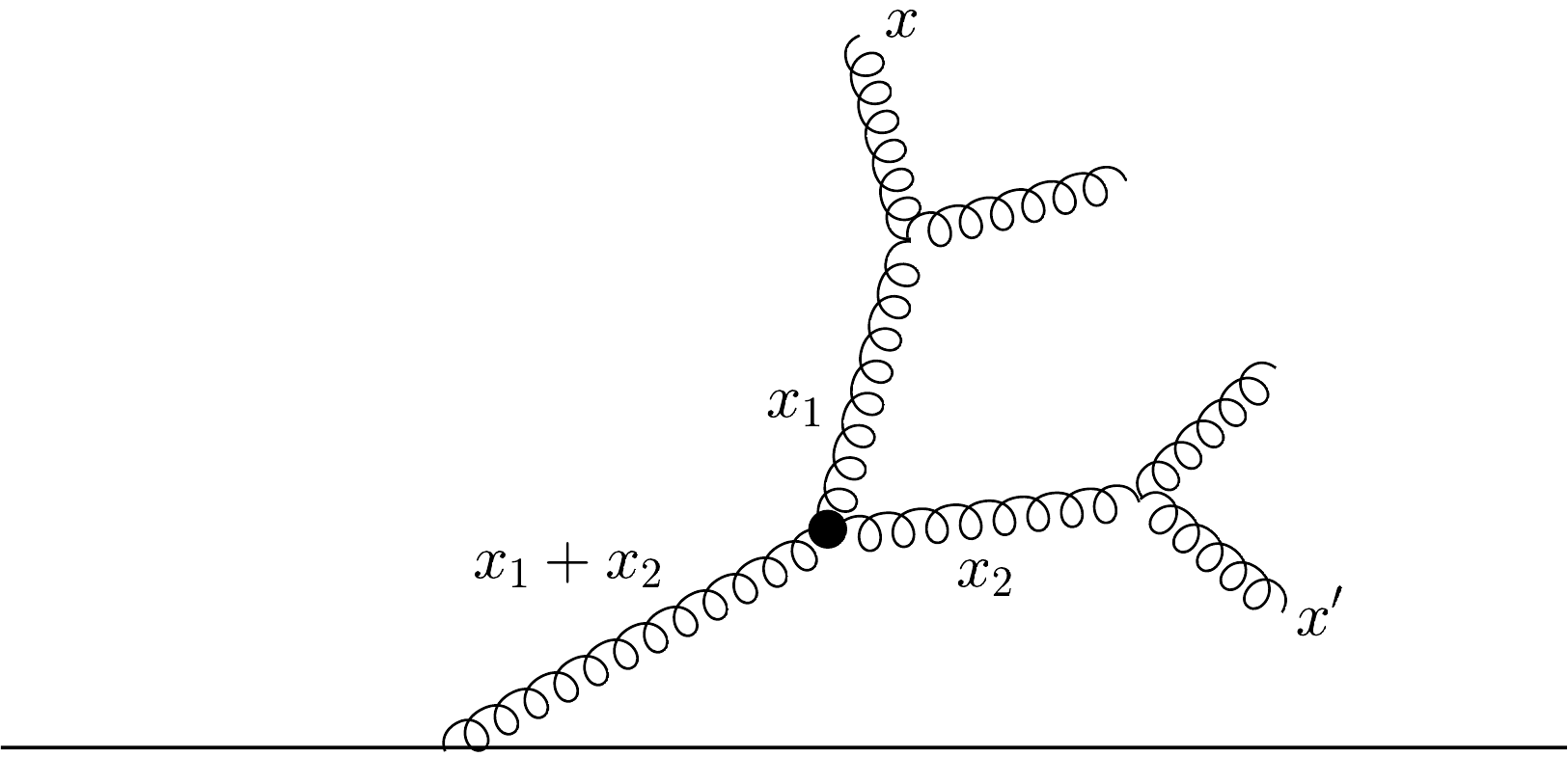} 
\end{center}
\end{minipage}
\begin{minipage}[b]{0.50\textwidth}
\begin{center}
\includegraphics[width=0.95\textwidth]{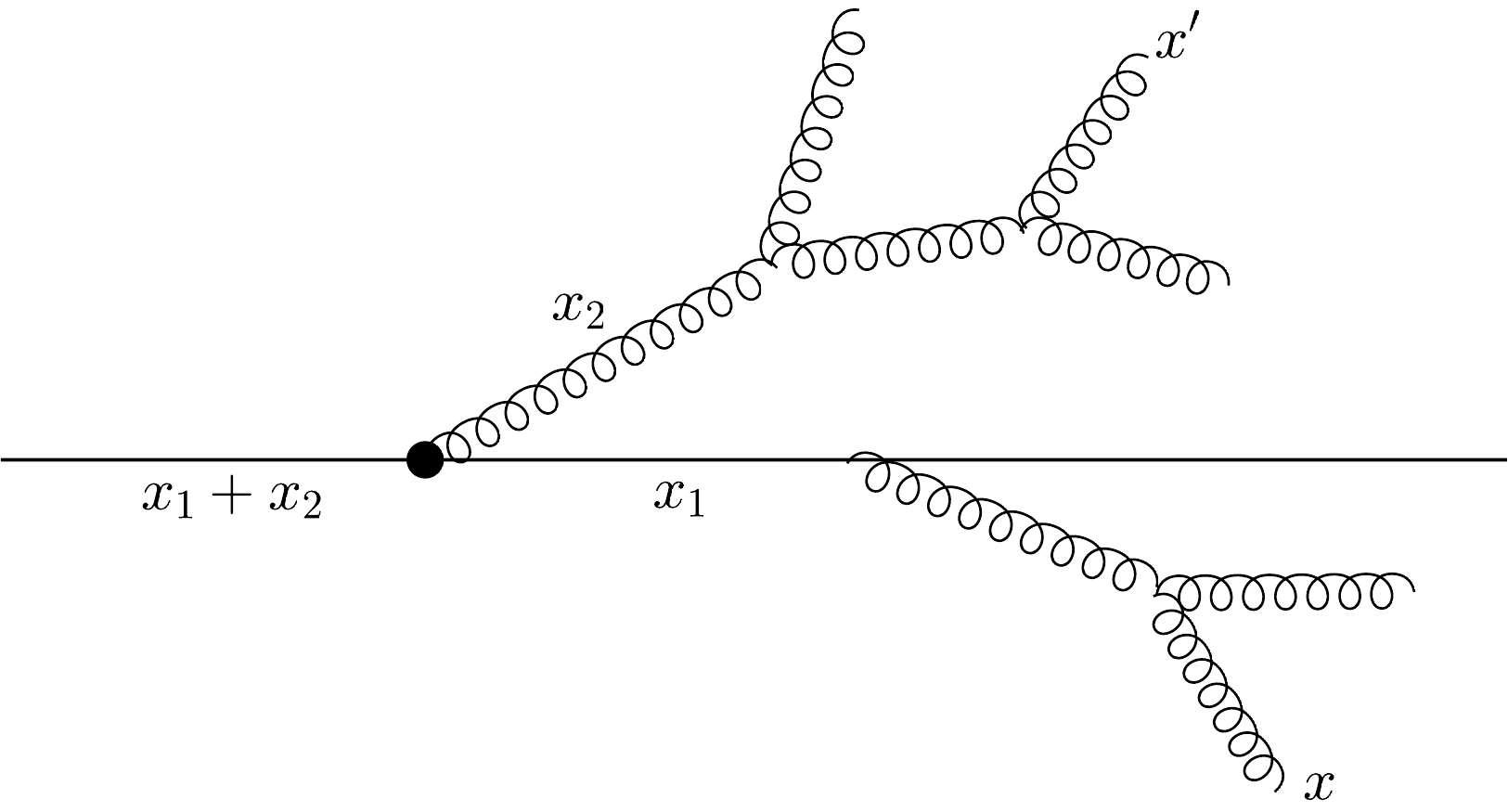}  
\end{center}
\end{minipage}
\begin{center}\caption{Pair correlations built via multiple branchings at 
relatively small times $\tau^2\ll 1$, when the LP still exists.
The continuous line represents the LP, while the wiggly lines are radiated gluons.
The measured gluons, with energy fractions $x,\,x'\ll 1$, have a last
common ancestor (LCA) with energy fraction $x_1+x_2$, whose splitting vertex 
is indicated by a blob. Left:  the LCA is a primary gluon with $x_1+x_2\ll 1$
and the two measured gluons belong to a same democratic cascade.
Right: the LCA is the LP itself  ($x_1+x_2\simeq 1$), which emits a primary
gluon with $x_2\ll 1$; in this case the two measured gluons belong to 
different mini-jets.
  \label{fig:correl}}
\end{center}
\end{figure}

The initial conditions for the above equations read $D(x,0)=\delta(x-1)$ and
$D^{(2)}(x,x',0)=0$. The respective solution $D(x,\tau)$ to \eqn{eqD}
acts as a Green's function for  \eqn{eqD2}: the solution to the latter can be written as
 \begin{align}\label{D2sol}
   D^{(2)}(x,x',\tau)\,=\,\int_0^\tau\rmd \tau'\int^1_{x}\frac{\rmd x_1}{x_1}\int^{1-x_1}_{x'}\frac{\rmd x_2}{x_2}\,
      D\bigg(\frac{x}{x_1},\frac{\tau-\tau'}{\sqrt{x_1}}\bigg) D\bigg(\frac{x'}{x_2},\frac{\tau-\tau'}{\sqrt{x_2}}\bigg)
\,S(x_1,x_2,\tau')\,, \end{align}
where $S(x_1,x_2,\tau')$ is a compact notation for the source term in the r.h.s. of \eqn{eqD2}. The physical
interpretation of \eqn{D2sol} is quite clear: at some intermediate time $\tau'$, a gluon with energy 
fraction $x_1+x_2$ splits into two daughter gluons with energy fractions $x_1$ and respectively $x_2$,
whose subsequent evolutions generate the mini-jets within which one measures the two final gluons $x$
and $x'$. Note that the parent gluon with energy $x_1+x_2$ is the {\em last common ancestor} (LCA) of the
two measured gluons $x$ and $x'$ : after this ancestor has split, the subsequent branchings leading to $x$
and $x'$ are completely disconnected from each other (see Fig.~\ref{fig:correl} for an
illustration).

\section{Event-by-event fluctuations in the energy loss at large angles}
\label{sec:var}
We are now prepared for an explicit calculation of the two quantities of primary interest
for us here: the average energy lost by the jet at large angles $\langle \mcal{E}(t)\rangle$
and its variance $\langle \mcal{E}^2(t)\rangle - \langle \mcal{E}(t)\rangle^2$. 
To obtain exact analytic results, we shall consider a slightly simplified version of the stochastic
process introduced above, as obtained by replacing the kernel ${\cal K}(z)$ in \eqn{Pdef}
with ${\cal K}_0(z)\equiv 1/{[z(1-z)]^{3/2}}$. This replacement is pretty harmless: 
the simplified kernel preserves the poles of the exact kernel at $z=0$ and $z=1$
and the respective residues, so it generates a very similar evolution. And indeed,
the numerical solutions to \eqn{eqD} using the exact kernel  \cite{Fister:2014zxa,Blaizot:2015jea}
are very similar, even quantitatively, to the exact analytic solution presented 
in Ref.~\cite{Blaizot:2013hx} for the simplified kernel ${\cal K}_0(z)$.

\subsection{The gluon spectrum and the average energy loss}
\label{sec:spec}

\begin{figure}[t]
	\centering
	\includegraphics[width=0.65\textwidth]{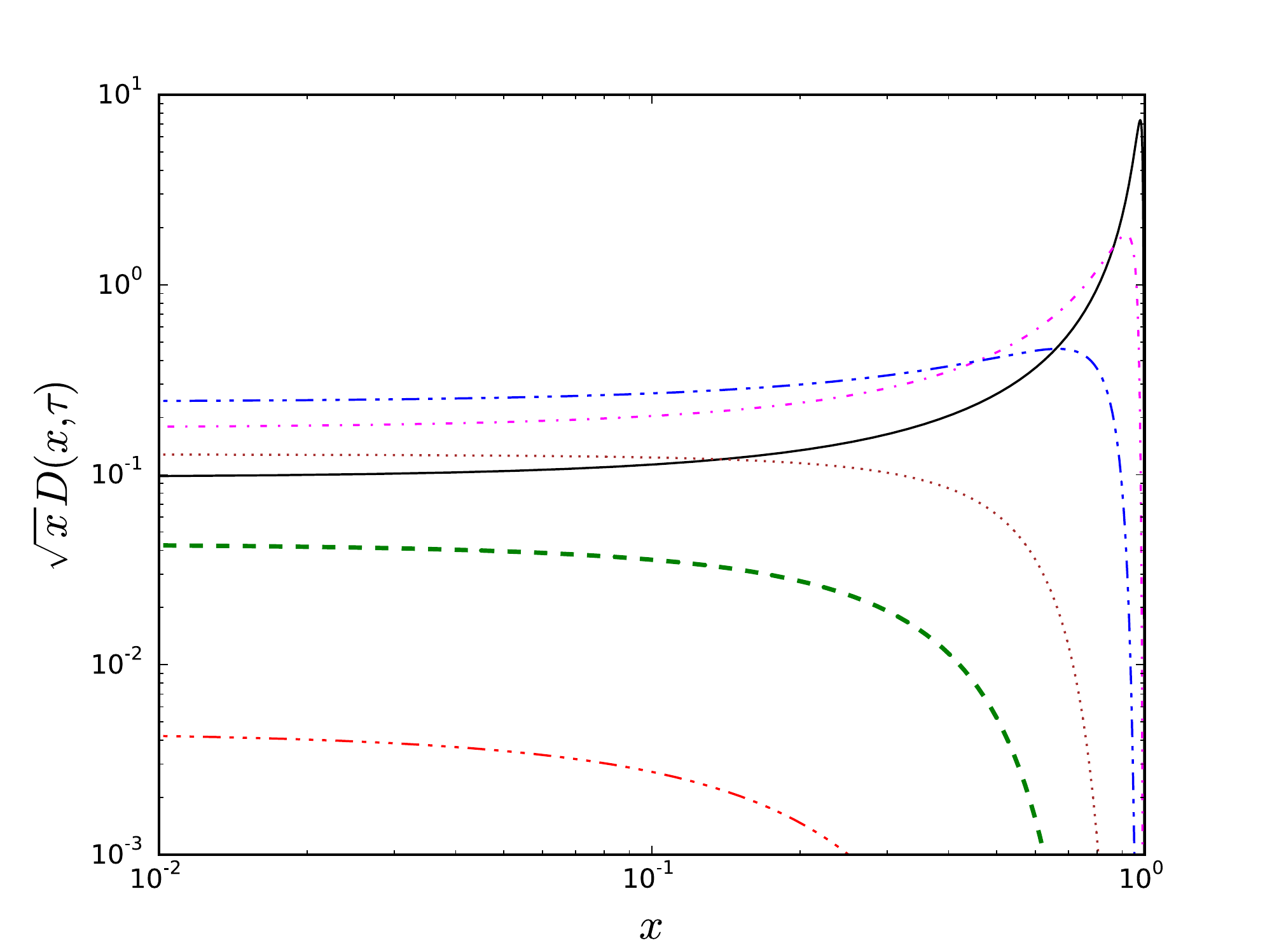}
		\caption{\sl Plot (in log-log scale) of $\sqrt{x} D(x,\tau)$, 
		with $D(x,\tau)$ given by Eq.~(\ref{Dexact}), as a function of $x$ for various values of $\tau$:
		solid (black): $\tau=0.1$; dashed--dotted (purple): $\tau=0.2$;
 dashed double--dotted (blue): $\tau=0.4$; dotted (brown): $\tau=0.75$;
  dashed (green): $\tau=1$; dashed triple--dotted (red) : $\tau=1.35$.
		}
		\label{fig:D}
\end{figure}

The solution to \eqn{eqD} with ${\cal K}(z) \to {\cal K}_0(z)$ and
the initial condition $D(x,0)=\delta(x-1)$ reads \cite{Blaizot:2013hx}
\beq\label{Dexact}
  D(x,\tau)\,=\,\frac{\tau}{\sqrt{x}(1-x)^{3/2}}\ \exp\left\{-\frac{\pi\tau^2}{1-x}\right\}\,.\eeq  
  This solution is illustrated in Fig.~\ref{fig:D}.
Its physical interpretation is important for what follows, so we shall now elaborate on it.
To that aim, let us first recall that 
\beq\label{tau}
\tau^2\,\equiv \,\frac{t^2}{\tbr^2(E)}\,=\,\frac{\obr(t)}{E}\,,
\eeq
where $\obr(t)=\abar^2\hat q t^2$ is the energy of the hardest emissions that
can occur with a probability of $\order{1}$ during a time $t$. The most interesting
regime for us here is the small-$\tau$ regime $\tau\ll 1$, which corresponds to the
typical experimental situation at the LHC: a very energetic LP with $E\gg\obr(L)$ or,
equivalently, a relatively small medium size $L\ll \tbr(E)$. We recall that $L$ is 
the total distance travelled by the LP through the medium and hence
the maximal value of $t$.

At small times, such that $\pi\tau^2\ll 1$, \eqn{Dexact} exhibits a pronounced 
peak just below $x=1$, which describes the leading particle. The position $x_p(\tau)$ 
of the peak (the local maximum of the function $D(x,\tau)$) represents the most
likely value of the energy of the LP at time $\tau$. Hence, the difference $1-x_p$ is the 
most likely value for its energy loss, that is, the energy loss by the LP in a 
typical\footnote{In this context, the distinction between the {\em typical} energy loss 
and the respective {\em average} quantity is indeed important: for the energy
loss by the leading particle, these two quantities are indeed different \cite{Baier:2001yt}; 
see  Sect.~\ref{sec:dis} below.} event. 
One easily finds $1-x_p \simeq (2\pi/3)\tau^2$, or,
in physical units, $\Delta E \equiv E(1-x_p)\sim  \obr(t)$, in agreement with 
the discussion in Sect.~\ref{sec:phys}: during a time $t$, the LP radiates with
probability of $\order{1}$ gluons with energies $\omega\lesssim \obr(t)$.
Its energy loss is dominated by the hardest among these emissions. 
The broadening of the LP peak, say, as measured by its width $\delta x$ at half the 
height of the peak, is itself of order $\pi\tau^2$. Clearly, this width $\delta x$ is a 
measure of the dispersion in the energy distribution of the LP. The fact that the
dispersion is comparable with the mean energy loss, $\delta x\sim 1-x_p \sim \tau^2$,
is an interesting observation, whose interpretation will be discussed in Sect.~\ref{sec:dis}.

Consider also the tail of the gluon spectrum at small $x$, which describes the soft
radiation. In view of the previous discussion, one might expect an accumulation of gluons
in the small-$x$ bins at $x\lesssim\pi\tau^2$, but this is actually not the case:
for  $x\ll 1$, \eqn{Dexact} exhibits the same power-law tail 
$D(x,\tau)\propto 1/\sqrt{x}$ 
as the BDMPSZ spectrum generated via a single emission by the LP:
\beq\label{Dsmallx}
  D(x,\tau)\,\simeq\,\frac{\tau}{\sqrt{x}}\,\rme^{-\pi\tau^2}\,\simeq\,\frac{\tau}{\sqrt{x}}\,,
  \eeq
with the second approximation valid at small times $\pi\tau^2\ll 1$. In other terms,
the spectrum at small $x$ is not modified by multiple branching. Mathematically,
this is possible because the power-law spectrum in
\eqn{Dsmallx} represents a fixed point\footnote{In the theory of
wave turbulence, this is known as a Kolmogorov-Zakharov fixed point 
\cite{KST,Nazarenko}.}  for the rate equation \eqref{eqD} \cite{Baier:2000sb,Blaizot:2013hx}:
the `gain' and `loss' terms precisely cancel each other. This condition is necessary
to ensure that the energy flux associated with multiple branching is independent of $x$
(`wave turbulence'); see \cite{Fister:2014zxa} for a calculation of this flux.

At later times $\pi\tau^2\gtrsim 1$, the LP peak disappears and the function
$D(x,\tau)$ has support at $x\lesssim 1/\pi\tau^2$. With increasing $\tau$, both the support
in $x$ and the strength of the spectrum at any point $x$ within this support 
are rapidly decreasing, meaning that the energy flows outside the spectrum.
To characterize this flow, we now compute the average energy fraction
carried by the whole parton cascade at time $\tau$. Using Eqs.~\eqref{avvep} and 
\eqref{Dexact}, one easily finds \cite{Blaizot:2013hx}
\beq\label{Xtau}
\langle X(\tau) \rangle = \int_{0}^1\rmd x\, D(x,\tau)=\rme^{-\pi\tau^2}\tau
\int_0^\infty\frac{\rmd u}{\sqrt{u}}\,\rme^{-\pi\tau^2 u}=\rme^{-\pi\tau^2}\,,
\eeq
where in performing the integral it was convenient to change variable as
$u\equiv x/(1-x)$.  The integral is controlled by $u\simeq 1/\pi\tau^2$, meaning
by $x\simeq 1/(1+\pi\tau^2)$. At small times $\pi\tau^2\ll 1$, this yields
$x\simeq 1-\pi\tau^2$, showing that, not surprisingly, most of the jet energy is 
carried by the LP.  What is more interesting though, is that the {\em missing} energy
at small times, namely 
\beq\label{veptau}
\langle \varepsilon(\tau)\rangle\,=\,1-\rme^{-\pi\tau^2}\,\simeq\,\pi\tau^2\,,
\eeq
or, in physical units (cf. \eqn{tau}),
\beq\label{Eloss}
\langle\mcal{E}(t)\rangle\,
\,=\,E\left[1-\rme^{-\pi\frac{\obr(t)}{E}}\right]\,\simeq\,\pi\obr(t)\,,
\eeq
is comparable with the energy radiated by the LP in a
typical event (recall the discussion after  \eqn{Dexact}). This is consistent
with the above observation that there is no energy accumulation in gluon spectrum at 
$x> 0$ : via successive democratic branchings, the primary gluons 
with $x\lesssim\tau^2$ transmit 
their energy to arbitrarily soft quanta, which formally accumulate at $x=0$.
(Physically, they thermalize and thus transmit their whole energy to the medium \cite{Iancu:2015uja}.)

\begin{figure}[t]
	\centering
	\includegraphics[width=0.65\textwidth]{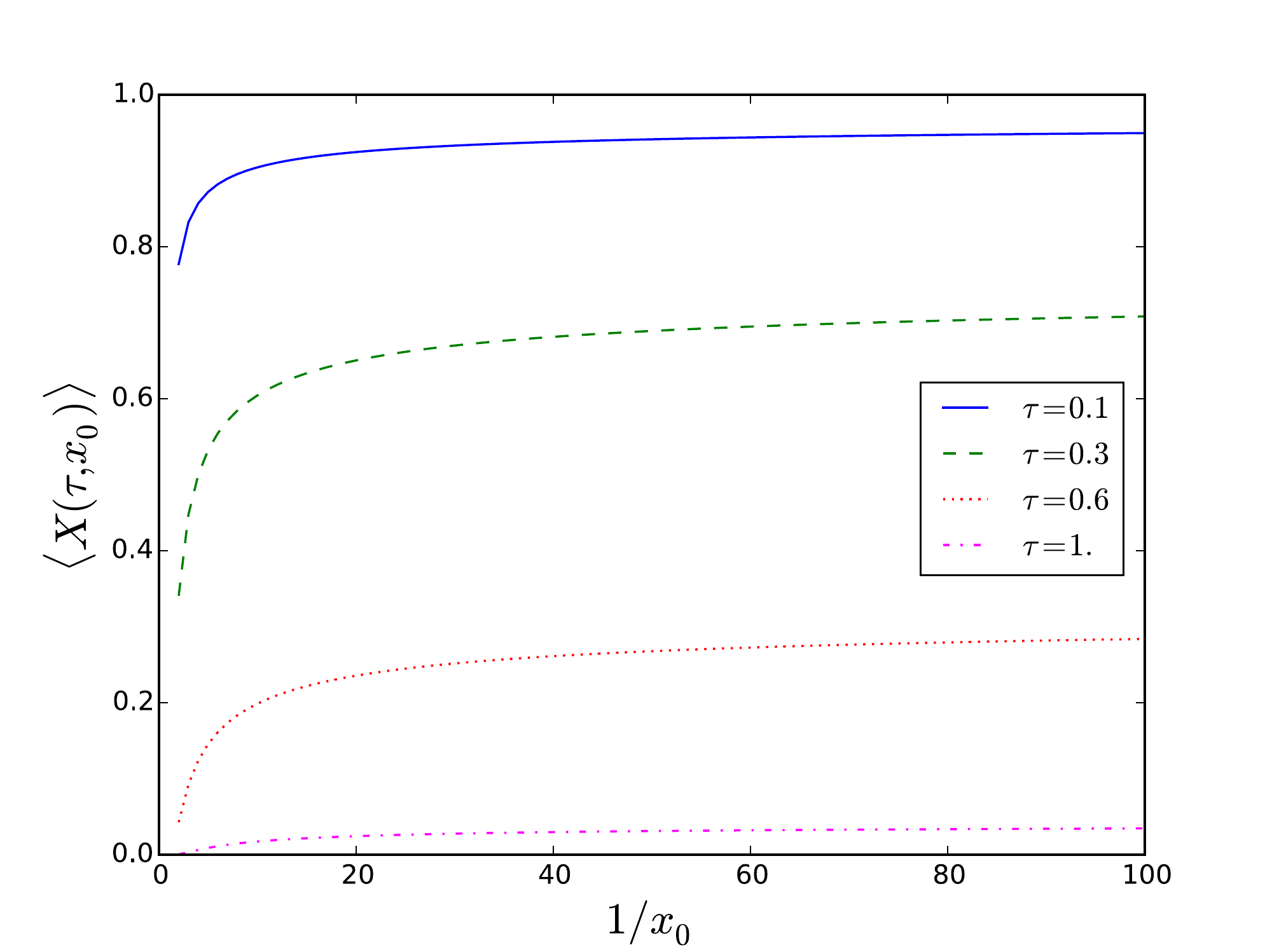}
		\caption{\sl The average energy fraction $\langle X(\tau,x_0)\rangle$
		carried by particles with $x\ge x_0$, as given by \eqn{Xtau0},
 is plotted as a function of $1/x_0$ for $x_0\le 0.5$
		 and various values of $\tau$:
		solid (blue): $\tau=0.1$; dashed (green): $\tau=0.3$;
 dotted (red): $\tau=0.6$; dashed--dotted (magenta): $\tau=1$.
		}
		\label{fig:X}
\end{figure}

It is also interesting to study the energy carried by the gluons in the spectrum
having an energy fraction larger than a given value $x_0\ll 1$. Via the relation
$\sin\theta_0=k_\perp/\omega_0$, with $\omega_0=x_0E$ and
$k_\perp^2\sim Q_s^2=\hat q L$, this  gives us an estimate of
the jet energy which propagates inside a cone with opening angle $\theta_0$. (Recall
the discussion around \eqn{Ex0}.) 
The corresponding generalization of \eqn{Xtau} reads 
\begin{align}\label{Xtau0}
\langle X (\tau, x_0) \rangle & = \int_{x_0}^1\rmd x\, D(x,\tau)=\rme^{-\pi\tau^2}\tau
\int_{u_0}^\infty\frac{\rmd u}{\sqrt{u}}\,\rme^{-\pi\tau^2 u}= \nonumber\\*[0.2cm]
&=\rme^{-\pi\tau^2}
\left[1-\text{erf}\left(\tau\sqrt{\frac{\pi x_0}{1-x_0}}\right)\right]
\,\simeq\,\rme^{-\pi\tau^2}\big(1-2\tau\sqrt{x_0}\big)
\,,
\end{align}
where $u_0\equiv x_0/(1-x_0)$ and we have introduced the error function
\beq\label{eq:erfc}
\text{erf}(a)\equiv \frac{2}{\sqrt{\pi }}\int _0^a\rmd z \,\rme^{-z^2}\,=\,
\frac{2}{\sqrt{\pi }}\left(a -\frac{a^3}{3} + \frac{a^5}{10} +\order{a^7}\right) \,.
\eeq
The last, approximate, equality in \eqn{Xtau0} holds when $\tau\sqrt{x_0}\ll 1$.

In Fig.~\ref{fig:X}, we exhibit $\langle X (\tau, x_0) \rangle$ as a 
function of $1/x_0$ for several interesting values of $\tau\le 1$. As manifest in this figure,
$\langle X (\tau, x_0) \rangle$ is essentially flat at $1/x_0\gg 1$. In other terms,
the total energy contained inside the jet increases only slowly when
increasing the jet angle $\theta_0\propto1/x_0$. This is easy to understand in the
present context: when $x_0\ll 1$,
the energy contained in the soft tail of the spectrum at $0<x< x_0$ is very small and 
truly negligible not only compared to the energy carried by the harder gluons 
with $x\gg x_0$, but also compared to the `missing' energy 
$\langle \varepsilon\rangle$ that has
accumulated at $x=0$. (In Fig.~\ref{fig:X}, this missing energy is represented by the 
`offset' $1-\langle X (\tau, x_0) \rangle$ of the various curves at $1/x_0\to\infty$.)
This discussion reenforces our previous argument, around \eqn{E0},
that the energy loss at large angles can be accurately computed by letting 
$x_0\to 0$ in  \eqn{Ex0}. In principle, one needs a non-zero cutoff $x_0
\propto 1/\theta_0$,
to control the angular distribution of the radiation. However, the values of $x_0$ 
corresponding to large angles $\theta_0\sim 1$ are quite small, $x_0\ll 1$, 
and for them the $x_0$-dependence of the energy distribution is truly negligible, 
as shown by Fig.~\ref{fig:X}. This ultimately reflects the fact that the energy loss
at large angles is controlled by the turbulent energy flow, which is independent
of $x$.

\subsection{The gluon pair density and the variance in the energy loss}
\label{sec:pair}

We now turn to our main new calculation, that of the variance $\sigma_\varepsilon^2(\tau)$ in 
the energy loss at large angles. As explained in Sect.~\ref{sec:eqs}, this involves the gluon pair 
density $D^{(2)}(x,x',\tau)$, for which an exact but formal solution has been written in \eqn{D2sol}.
Remarkably, it turns out that for the simplified kernel  ${\cal K}_0(z)$ 
(i.e., for the gluon spectrum $D(x,\tau)$ in  \eqn{Dexact}), the integrals in \eqn{D2sol} can be computed
exactly, with the following result (see App.~\ref{app:D2} for details):
\begin{align}\label{D2exact}
   D^{(2)}(x,x',\tau)\,=\,\frac{1}{2\pi}\frac{1}{\sqrt{x x' (1-x-x')}}\left[\rme^{-\frac{\pi\tau^2}{1-x-x'}}
   -\rme^{-\frac{4\pi\tau^2}{1-x-x'}}\right].
   \end{align}

To get more insight into this result, consider the particular case where $\pi\tau^2\ll1$ (that is,
the LP has no time to undergo a democratic branching) and $x,\,x'\ll 1$.
In this limit, \eqn{D2exact} reduces to 
\begin{align}\label{D2small}
   D^{(2)}(x,x',\tau)\,\simeq\,\frac{3}{2}\frac{\tau^2}{\sqrt{x x'}}\,\simeq\,\frac{3}{2}\,D(x,\tau) D(x',\tau)\,,
  \end{align}  
where we have also used \eqn{Dsmallx}. This result is truly remarkable in several respects.

\begin{figure}[t]
	\centering
	\includegraphics[width=0.9\textwidth]{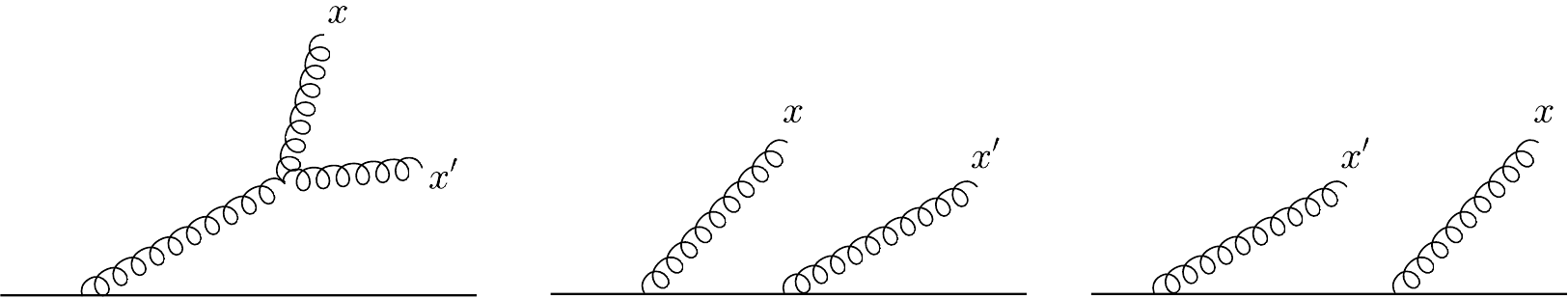}
		\caption{\sl Processes contributing to the production of 2 soft gluons
		to lowest order in perturbation theory.
		}
		\label{fig:2g}
\end{figure}

First, \eqn{D2small} coincides with the respective prediction of lowest-order perturbation
theory\footnote{By `perturbation theory'  we more precisely mean
the expansion in the number of branchings or, equivalently, 
in powers of $\tau$. For the correlations of interest, this is tantamount to solving Eqs.~\eqref{eqD}
and \eqn{eqD2} via iterations.}, that is, the 3 graphs in Fig.~\ref{fig:2g} which describe the emission of only 2 
soft gluons. (Indeed, one needs at least 2 branchings in order to produce 2 soft gluons in the final state.)
These 3 graphs can be easily evaluated and one finds that they give identical contributions; 
in particular, the overall factor $1/2$ comes in each of them from integrating over the branching 
time $\tau'$ of the last common ancestor (LCA): $\int_0^\tau\rmd\tau'\,\tau'=\tau^2/2$.

A priori, this perturbative estimate is expected to be correct for sufficiently hard emissions,
$x,\,x'\gg\tau^2$, but the result in \eqn{Dsmallx} also applies to {\em very} soft gluons, with
$x,\,x'\lesssim\tau^2$, for which the effects of multiple branchings are known to be
important --- actually, non-perturbative. This observation extends to the gluon pair density
an important property that we noted in
Sect.~\ref{sec:spec} in relation with the gluon spectrum (recall the discussion after \eqn{Dsmallx}):
the fact that the gluon distribution produced by the branching process at small $x$ is formally 
insensitive to multiple branching, since controlled by the turbulent fixed point.
Of course, the effects of multiple branching become visible when looking at the structure of
the leading-particle peak (see e.g. \eqn{D2exact} for $x+x'\simeq 1$) and also at {\em global}
properties, like the total energy enclosed in the spectrum. But, at least for small times $\tau\ll 1$
(so long as the LP still exists), they are not manifest in the gluon distribution at small $x$.

The physical interpretation of the result in \eqn{Dsmallx} for generic values  $x,\,x'\ll 1$
can be inferred from the explicit calculation of the integrals in
\eqn{D2sol}, in App.~\ref{app:D2}. This involves
3 types of processes which are topologically similar to those depicted in Fig.~\ref{fig:2g}
--- the latter must be simply dressed by multiple branchings. 

In one of
these processes, the two measured gluons $x$ and $x'$ belong to a same mini-jet
(see Fig.~\ref{fig:correl} left). That is,
their LCA, which has energy fraction $x_1+x_2$ in  \eqn{D2sol},
is itself soft ($x_1+x_2\ll 1$), but such that $x_1-x\sim \tau^2$ and similarly $x_2-x'\sim\tau^2$. 
These conditions ensure that the gluon $x$ is found with probability of $\order{1}$ in the cascade 
generated by the gluon $x_1$  
during a time interval $\tau-\tau'\sim\tau$, and similarly for the pair of gluons $x'$ and $x_2$. 
If the measured gluon, say $x$, is relatively hard, such that $x\gg\tau^2$ (with $x\ll 1$ though),
then $x_1\simeq x$, meaning that the energy fractions $x$ and $x_1$ refer to a {\em same} gluon, 
which emits some soft radiation in passing from $x_1$ to $x$. If on the other hand 
$x\ll\tau^2$, then $x_1\sim\tau^2\gg x$, so the measured gluon $x$ is one of the soft gluons in 
the cascade generated by $x_1$ via democratic branchings. A similar discussion applies
to $x'$ and $x_2$.

In the 2 other processes, the LCA is the LP ($x_1+x_2\simeq 1$), so
the two measured gluons belong to different mini-jets. E.g. the LP emits a 
primary gluon $x_2$ with $x_2-x'\sim\tau^2$, which then generates
a democratic cascade which includes the final gluon $x'$. The other final gluon, 
with energy $x$, is found in the mini-jet produced by another primary gluon,
which is later radiated by the LP
(see Fig.~\ref{fig:correl} right).

The difference $D^{(2)}(x,x',\tau)-D(x,\tau) D(x',\tau)$ (the `factorial cumulant') is a measure of the correlations
in the gluon distribution. We shall later argue, 
in Sect.~\ref{sec:dis} and in Sect.~\ref{sec:number}, that the successive emissions of primary gluons 
are quasi-independent from each other. This in turn implies that the net correlation
$D^{(2)}(x,x',\tau)-D(x,\tau) D(x',\tau)\simeq (1/2)D(x,\tau) D(x',\tau)$ is to be associated with
the first process described above, cf. Fig.~\ref{fig:correl} left, where the two measured gluons 
belong to a same mini-jet.

\begin{figure}[t]
\begin{center}
\centerline{
\includegraphics[width=0.53\textwidth]{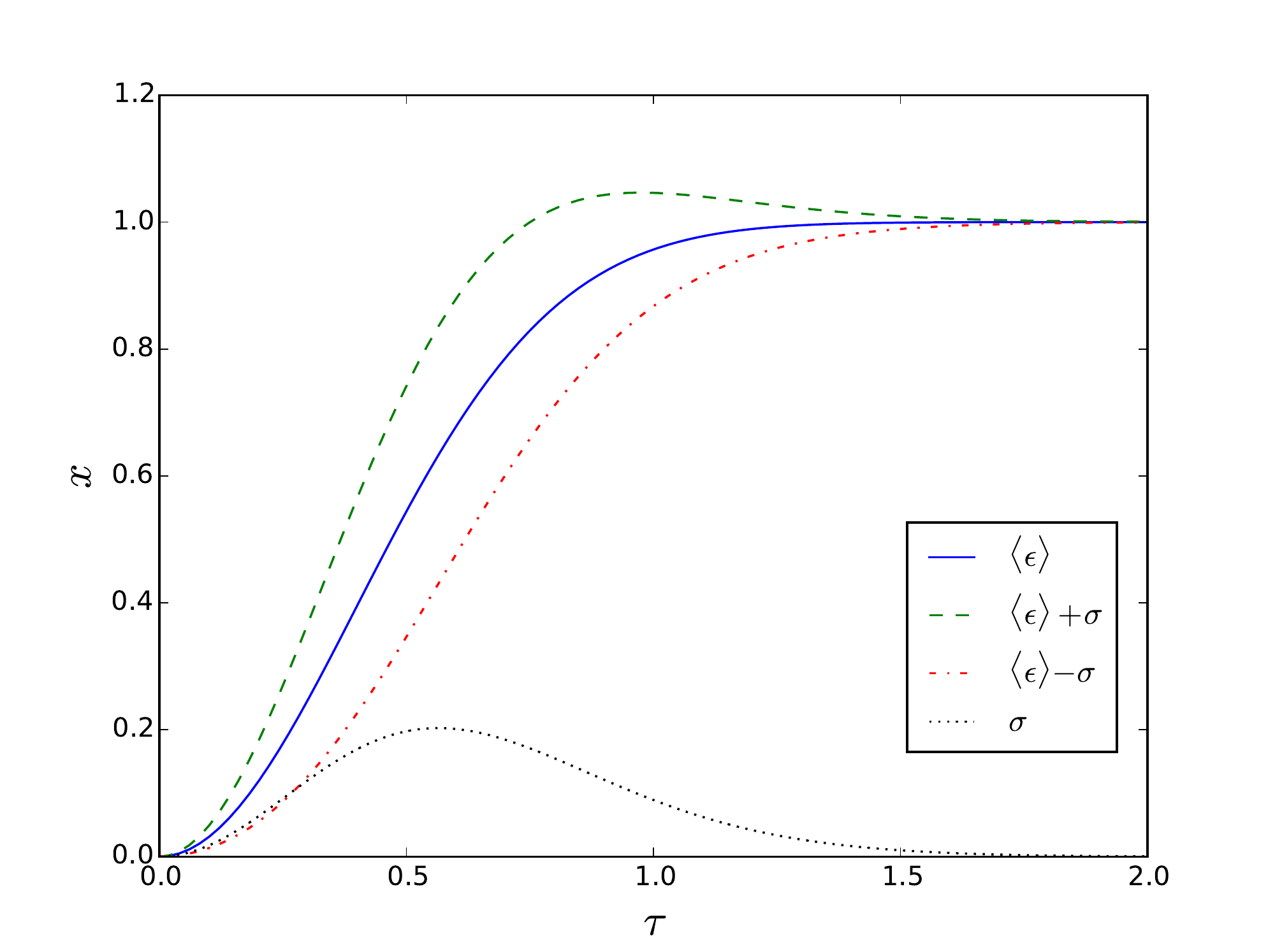}
\includegraphics[width=0.53\textwidth]{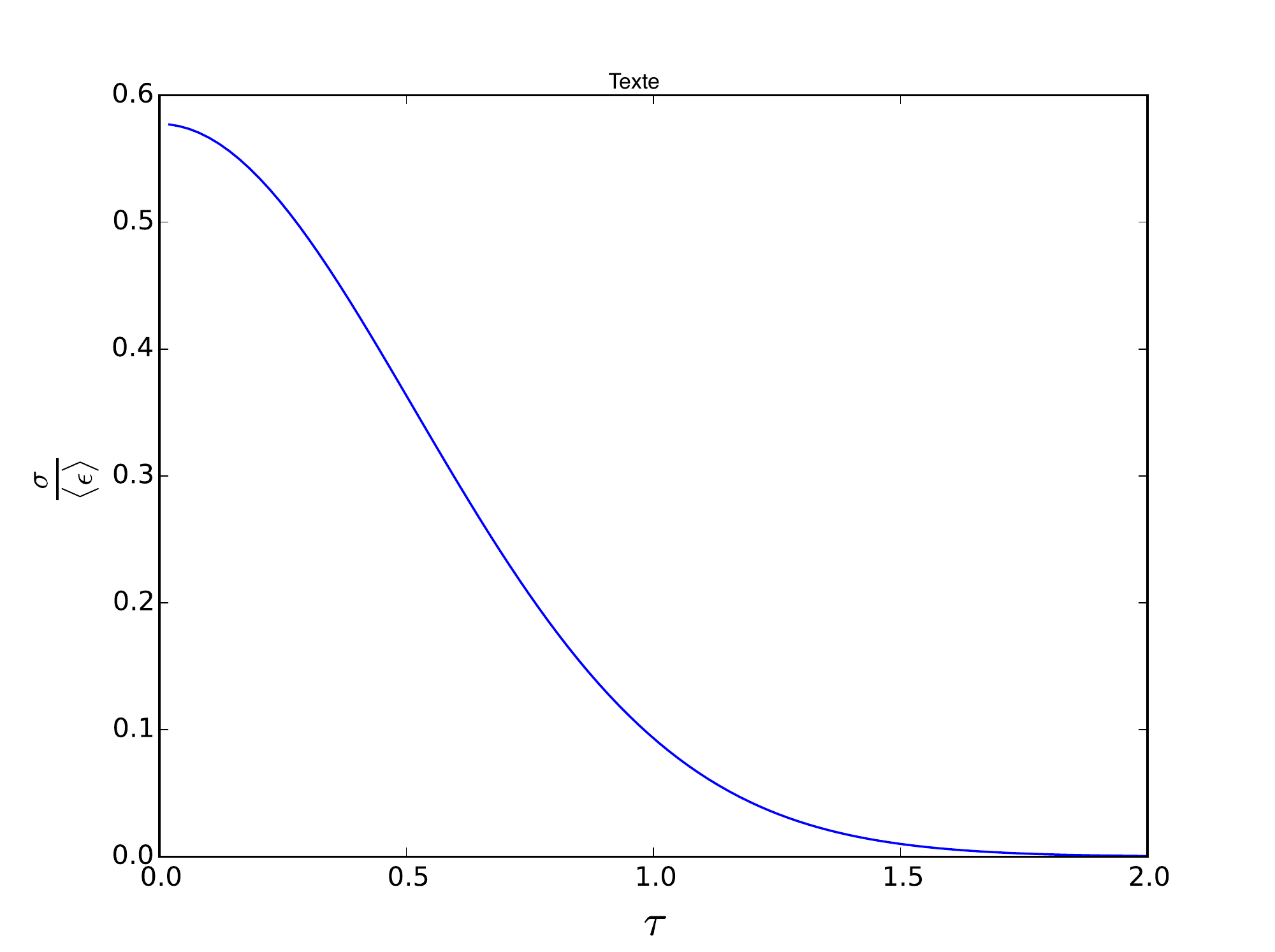}}
\caption{Left:  the time evolutions of the average (fractional) energy loss at large angles 
$\langle \varepsilon(\tau)\rangle$,
\eqn{veptau}, of the respective dispersion $\sigma_\varepsilon(\tau)$,
cf. \eqn{VAR}, and of the typical range $\langle \varepsilon\rangle- \sigma_\varepsilon\le 
\varepsilon\le \langle \varepsilon\rangle+ \sigma_\varepsilon$ for event-by-event
fluctuations in the random variable $\vep$. Right: the time evolution of
the dispersion-to-average ratio $\sigma_\varepsilon(\tau)/\langle \varepsilon(\tau)\rangle$. 
\label{fig:sigma}}
\end{center}
\end{figure}

Using \eqn{D2exact}, it is straightforward to compute the  first piece in the r.h.s. of \eqn{X2}
for $\langle X^2(\tau)\rangle$ (see App.~\ref{app:X2} for details): 
\begin{align}\label{X2A}
\int_{0}^1\rmd x \int_{0}^1\rmd x'\, D^{(2)}(x,x',\tau)
&=\pi\tau\big[1+ \text{erf}(\sqrt{\pi}\tau)-2 \text{erf}(2\sqrt{\pi}\tau)\big]
-\,\rme^{-4\pi\tau^2}+\,\rme^{-\pi\tau^2}
\nonumber\\*[0.2cm]
&=
\pi\tau-3\pi\tau^2+\frac{5}{2}(\pi\tau^2)^2-\frac{21}{10}(\pi\tau^2)^3
+\mathcal{O}(\tau^8)\,,
\end{align}
where we have used \eqn{eq:erfc}.
The second piece in \eqn{X2} is easily computed as (see App.~\ref{app:X2})
\begin{align}\label{X2B}
\int_{0}^1\rmd x\, x\,D(x,t)&\,=\,
\rme^{-\pi\tau^2} -\pi\tau\left[1-\text{erf}\left(\sqrt{\pi}\tau\right)\right]
\nonumber\\*[0.2cm]
&= 1 -\pi\tau + \pi\tau^2 -\frac{1}{6}(\pi\tau^2)^2 + \frac{1}{30}(\pi\tau^2)^3 + \order{\tau^8}
\end{align}
Note that  the terms linear in $\tau$ in the above small-time expansions cancel between the
two contributions to $\langle X^2(\tau)\rangle$. After also subtracting the square 
$\langle X(\tau)\rangle^2$ of the average value, we obtain the variance:
\begin{align}\label{VAR}
\sigma_\varepsilon^2(\tau)&=
2\pi\tau\big[\text{erf}(\sqrt{\pi}\tau)-\text{erf}(2\sqrt{\pi}\tau)\big]+2\rme^{-\pi\tau^2}
-\,\rme^{-4\pi\tau^2}-\rme^{-2\pi\tau^2}
\nonumber\\*[0.2cm]
&=\frac{1}{3}\pi^2\tau^4-\,\frac{11}{15}\pi^3\tau^6+\mathcal{O}(\tau^8)\,.
\end{align}
Remarkably, the terms of $\order{\tau^2}$ have cancelled between
$\langle X^2(\tau)\rangle$ and $\langle X(\tau)\rangle^2$, so the
 net contribution to $\sigma_\varepsilon^2(\tau)$ at small time is {\em quartic} in $\tau$.
This result is very interesting: it shows that, in the physically interesting regime at  $\pi\tau^2\ll 1$,
the dispersion $\sigma_\varepsilon(\tau)$ in the energy loss at large angles is 
comparable with the respective expectation value:
$\sigma_\varepsilon(\tau)\sim \langle \varepsilon(\tau)\rangle\sim \tau^2$. A
physical interpretation of this result will be presented in the next subsection.
For latter purposes, we notice that, to order $\tau^4$, \eqn{VAR} can be equivalently 
rewritten as 
$\langle \varepsilon^2(\tau)\rangle\simeq (4/3)\langle \varepsilon(\tau)\rangle^2$.

For larger times $\pi\tau^2\gtrsim1$, on the other hand, the variance rapidly vanishes,
\begin{align}\label{VARas}
\sigma_\varepsilon^2(\tau)\,\simeq \,\frac{1}{\pi\tau^2}\,\rme^{-\pi\tau^2}\bigg[
1 +\order{\frac{1}{\pi\tau^2}}\bigg]\,,\end{align}
as is should be expected: at large times, the {\em whole} energy is lost towards the `condensate'
(meaning, towards the medium), so the fluctuations vanish.

The results obtained in this section are illustrated in Fig.~\ref{fig:sigma}. As manifest
there, the event-by-event fluctuations in the energy loss at large angles are 
numerically important (comparable to the respective average value, albeit somewhat smaller) 
for all times $\tau < 1$, i.e. $t<\tbr(E)$. Their relative strength, as measured by the ratio
$\sigma_\varepsilon(\tau)/\langle \varepsilon(\tau)\rangle$, is particularly large at 
$\tau\lesssim 0.5$, which is the interesting regime for  the phenomenology
of high-energy jets at the LHC.
For instance, for a realistic medium size $L=4$\,fm and a LP with $E=100$\,GeV,
one finds (using $\hat q=1$\,GeV$^2$/fm and $\abar=0.3$) a value
$\tau=L/\tbr(E)\simeq 0.3$, for which the fluctuation to average ratio is
$\sigma_\varepsilon/\langle \varepsilon\rangle\simeq 0.5$.

\subsection{Physical discussion}
\label{sec:dis}

An important result of the previous analysis is that, in the kinematical
range of interest for di-jets at the LHC, the event-by-event fluctuations in the energy loss
at large angles are comparable to the respective average value. 
Specifically (cf. Eqs.~\eqref{VAR} and \eqref{Eloss}, that we here
rewrite in physical units),
\beq
\label{dispersion}
\sigma_{\mcal{E}}(t)\,=\,\frac{1}{\sqrt{3}}\,\langle\mcal{E}(t)\rangle\,=\,
\frac{\pi}{\sqrt{3}}\,
\obr(t)\,\sim\, 
\abar^2\hat q t^2\qquad\mbox{when\quad $t\ll \tbr(E)$}
\,.\eeq
To understand this result, one needs to identify the physical origin of the various terms which
occur in the small-$\tau$ expansion in  \eqn{X2A}, notably the quartic term --- the only one to survive
in the final result for the variance, \eqn{VAR}. To that purpose, one should recall that the double integral
$\int_{0}^1\rmd x \int_{0}^1\rmd x'\, D^{(2)}(x,x',\tau)$ measures the total energy squared carried by
pairs of gluons from the jet; hence, it is naturally biased towards large values for $x$ and $x'$.

\begin{figure}[t]
\begin{minipage}[b]{0.50\textwidth}
\begin{center}
\hspace*{-1.cm}
\includegraphics[width=0.88\textwidth]{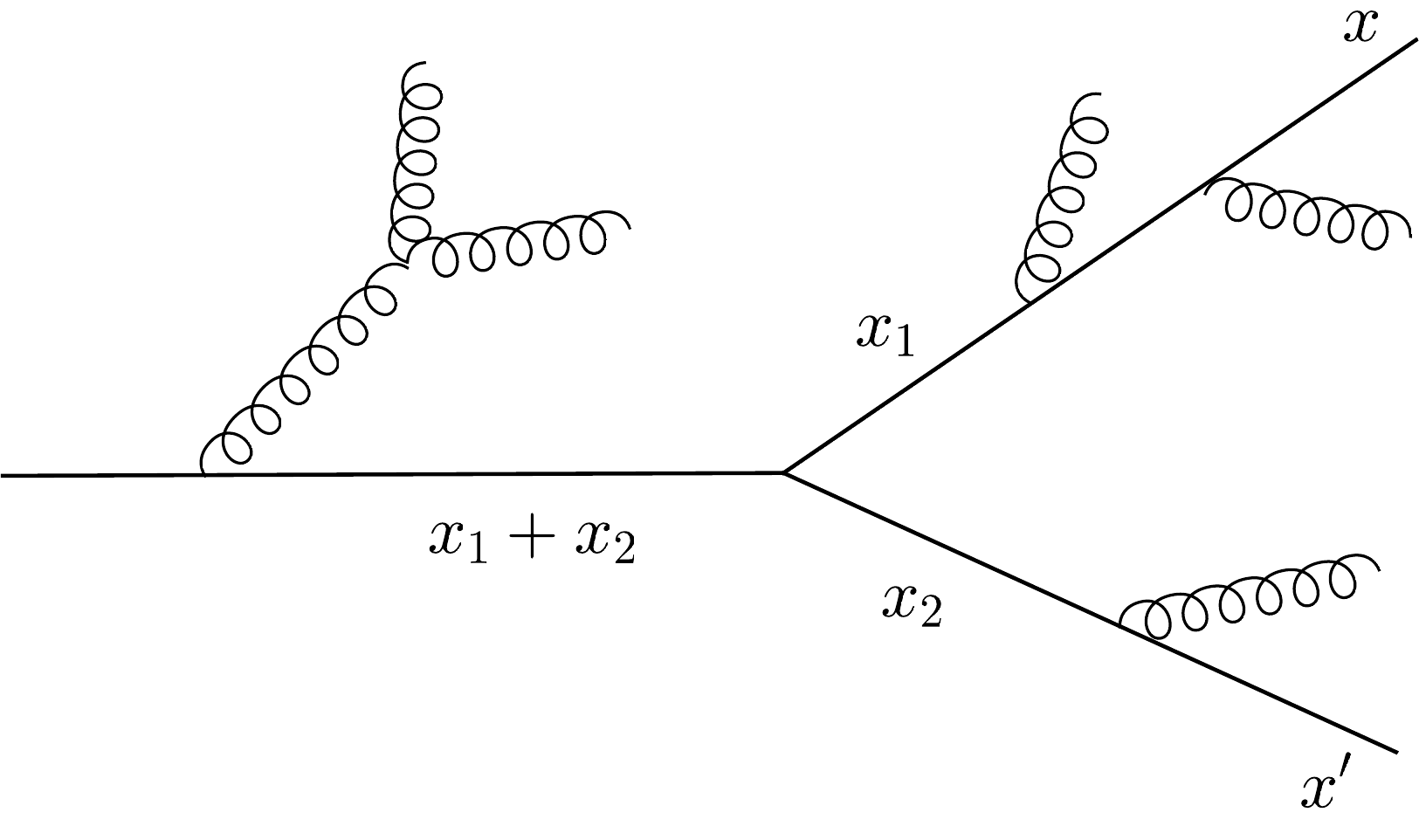} \quad 
\end{center}
\end{minipage}
\begin{minipage}[b]{0.50\textwidth}
\begin{center}
\vspace*{-1.cm}
\includegraphics[width=1.0\textwidth]{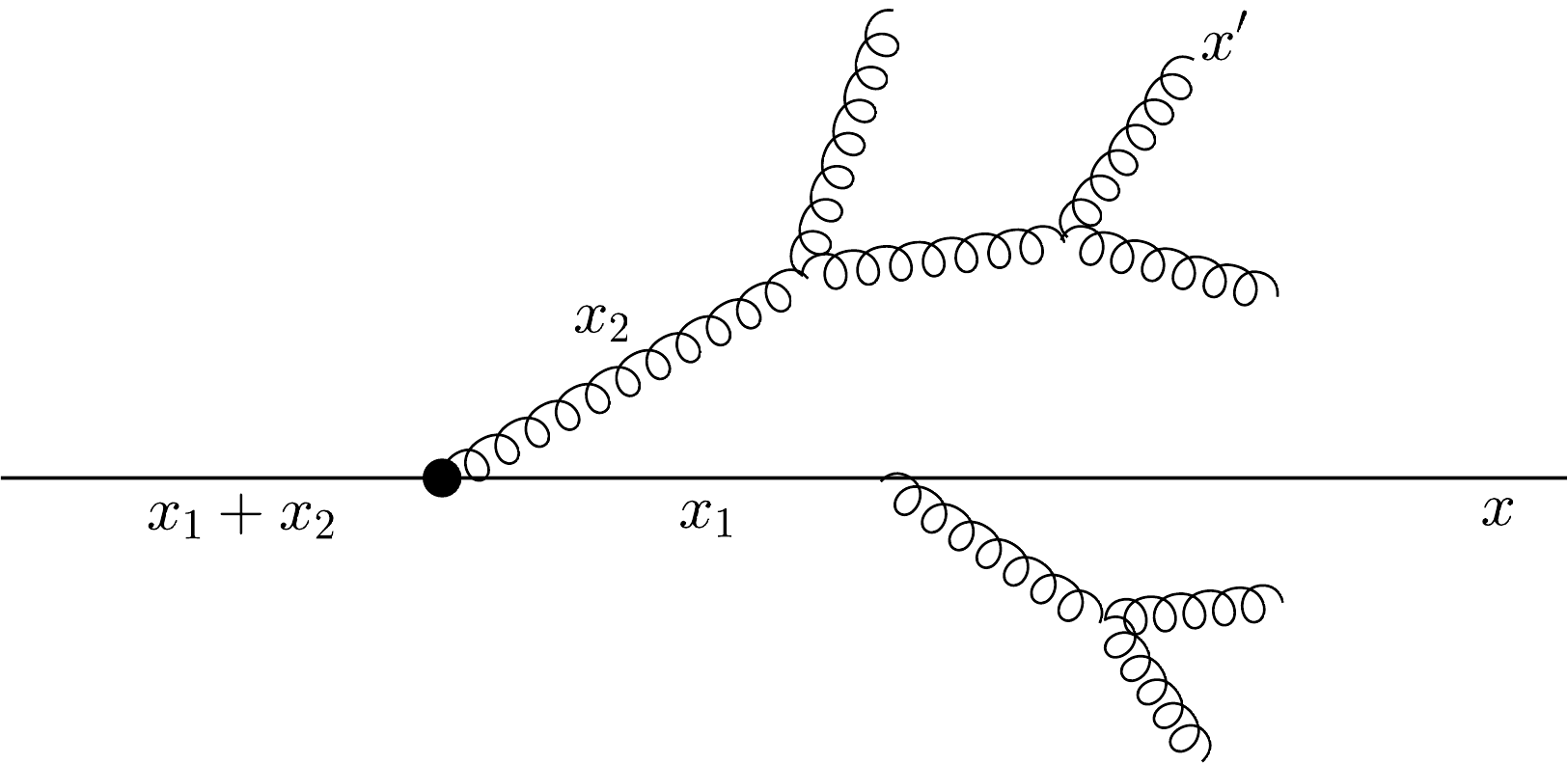}  
\end{center}
\end{minipage}
\begin{center}\caption{Left: the hard democratic splitting which produces the leading-order contribution,
of $\order{\tau}$, to the small-$\tau$ expansion in  \eqn{X2A}; the continuous lines represent
hard gluons with $x_1\simeq x_2\simeq 1/2$; the wiggly lines are soft gluons with
$x\lesssim \tau^2$.   Right: a branching sequence responsible for the contribution of $\order{\tau^2}$.
   \label{fig:var}}
\end{center}
\end{figure}

By inspection of the calculations in App.~\ref{app:X2}, it is rather easy to trace back the term linear
in $\tau$, which is the dominant term at small $\tau$. This is generated by a truly democratic
branching of the LP: the latter splits into two 
daughter gluons with $x_1\simeq x_2\simeq 1/2$, which are eventually measured: 
$x\simeq x_1$ and $x'\simeq x_2$ (see Fig.~\ref{fig:var} left). 
Such a hard democratic branching has a low probability, of order $\tau=t/\tbr(E)\ll 1$, but it generates a pair 
of gluons with the highest possible energies,
hence it dominates the double integral in  \eqn{X2A}. The order of magnitude
of this contribution is set by the probability $\sim\order{\tau}$ for the hard branching.

Furthermore, the term quadratic in $\tau$ comes from processes in which the LP radiates
with probability of $\order{1}$
a primary gluon with energy fraction $x_2\sim\tau^2$ (i.e. with energy $\omega_2\sim\obr(t)$),
which then initiates a democratic cascade (with probability of $\order{1}$, once again).
In this case the measured gluons are the LP itself (with energy fraction $x\sim x_1\sim 1$) and
one of the relatively hard gluons in the democratic cascade generated by $x_2$, 
namely a gluon having $x'\sim x_2\sim\tau^2$.
Such a branching sequence is illustrated in Fig.~\ref{fig:var} right. (The symmetric process, 
where $x_1$ and $x_2$ exchange their roles, gives an identical contribution.)
The overall order of magnitude is now set by the total energy $\sim\tau^2$ 
carried by the softest measured gluon $x'$.

Finally, the term of $\order{\tau^4}$, which is the most interesting one for our purposes, receives
contributions from the processes previously discussed in relation with  \eqn{D2small} and illustrated
in Fig.~\ref{fig:correl}.  In this case, the two measured gluons have energy fractions
$x\sim x'\sim\tau^2$\,; that is, they are soft relative to the LP, yet they are among the hardest
gluons that can be emitted with a probability of $\order{1}$ during a time $t$. (The order of 
magnitude of the respective contribution is then set by the product $x x'\sim \tau^4$.)
This in particular
implies that all the {\em primary} gluons that enter the processes shown in Fig.~\ref{fig:correl} (e.g.,
the gluon with energy fraction $x_1+x_2$ in Fig.~\ref{fig:correl} left) have themselves
energy fractions of order $\tau^2$.


We see that the dominant contribution, of $\order{\tau^2}$, to the dispersion $\sigma_\varepsilon(\tau)$
at small $\tau$ has the same physical origin as the respective contribution to
the average energy loss $\langle \varepsilon(\tau)\rangle$: the relatively hard 
primary emissions with $x\sim\tau^2$, or  $\omega\sim\obr(t)$. This observation naturally explains the 
result in \eqn{dispersion}, via the following chain of arguments:
 {\sf (a)}  the average number of such primary gluons 
is parametrically of order one, {\sf (b)} their emissions are random and quasi-independent from each other,
hence {\sf (c)} the fluctuations in their number are of order one as well. 

\comment{Point {\sf (a)} has been anticipated in the physical discussion in Sect.~\ref{sec:phys}.
It has been further substantiated by the calculation of the average energy loss, cf.
\eqn{Eloss}, and of its dispersion, cf. \eqn{VAR} and the physical discussion at the
end of Sect.~\ref{sec:var}.
As explained there, democratic branching is the most efficient mechanism for 
energy transfer from the LP towards soft quanta. Hence, the energy
lost by the jet at large angles is controlled by the hardest primary gluons which
can undergo democratic branchings during a time $t$ --- those having energies 
$\omega\sim\obr(t)$.}

Point {\sf (a)} follows from the very definition of the scale $\obr(t)$, as the 
energy of the gluon emissions 
which occur with probability of order one during the time $t$. 
This is further corroborated by the fact that both the average energy loss, 
\eqn{Eloss}, and the typical energy loss by the LP, cf. the discussion
after \eqn{Dexact},
can be expressed as $\obr(t)$ times a number which is parametrically of $\order{1}$.

Point {\sf (b)} can be understood as follows: the successive emissions of primary gluons 
with $\omega\sim\obr(t)$ are uncorrelated with each other, since they cannot overlap in time.
Indeed, these emissions are distributed in time during the interval $t$, 
the formation time for each of them $\tf(\obr(t))\sim \abar t$ is parametrically shorter than
$t$, and the overall number of such emissions
is of $\order{1}$. There is of course a correlation introduced by
energy conservation, but this is rather weak, since the total energy
carried by these emissions, of order $\obr(t)$, is much smaller than the available energy,
which is $E$. 

Finally, point {\sf (c)} is an immediate consequence of the two previous points. 
Together with point {\sf (a)}, it ultimately implies that both the average 
energy loss at large angles and its typical fluctuations should be of order $\obr(t)$, 
in agreement with \eqn{dispersion}.

The above arguments also explain the gross structure of the 
LP peak in the gluon spectrum \eqref{Dexact}, i.e. the fact that the width 
$\delta x\sim \tau^2$ of this peak is of the same order as the {\em typical} energy
loss by the LP, as measured by the shift $1-x_p \sim \tau^2$ 
in the position of its maximum.
The emphasis on the {\em typical} energy loss by the LP in the above discussion
is indeed important, since this quantity is different from the respective {\em average} 
quantity $\langle \Delta E\rangle$: the former is associated with relatively soft primary 
emissions, with energies $\omega\lesssim\obr(t)$, which occur event-by-event, whereas
the latter is controlled by the hardest possible emissions, with energy
$\sim\omega_c(t)\equiv \hat q t^2$  \cite{Baier:1996kr,Zakharov:1996fv}. 
Such hard emissions are rare events, which
occur with a probability of $\order{\abar}$ (cf. \eqn{Pdeltat}), but they take 
away a large amount of energy and thus give the dominant contribution 
$\langle \Delta E\rangle\sim \abar\hat q t^2$ to the average energy lost by the LP.


The previous discussion also suggests that the emission of soft primary gluons should 
be a Poissonian  process (see also Ref.~\cite{Baier:2001yt} for a similar argument). 
That is, the probability to radiate $n$ gluons during a time $t$, such that
each gluon is soft relative to its emitter (the LP), but harder than some `infrared'
scale $\omega_0$, is of the Poisson type, with the splitting rate given by \eqn{Pdeltat}:
\beq\label{Poisson}
\mcal{P}(n, t;\omega_0)\,=\,\frac{\big(\gamma(\omega_0) t\big)^n\,\rme^{- \gamma(\omega_0) t}}
{n!}\,,\qquad \gamma(\omega_0)\equiv \frac{2{\abar}}{\tf(\omega_0)}
 \,=\,2{\abar}\sqrt{\frac {\hat q}{\omega_0}}\,,
\eeq
This distribution predicts
that the average number of primary emissions and its variance are equal to each other,
which truly means that there are no correlations: $\langle n(n-1)\rangle=
\langle n\rangle^2$. Specifically,
\beq\label{nPoisson}
\langle n(t,\omega_0)\rangle\,=\,
\langle n^2(t,\omega_0)\rangle- \langle n(t,\omega_0)\rangle^2 \,=\,
\gamma(\omega_0) t\,=\,2\left[\frac{\obr(t)}{\omega_0}\right]^{1/2}\,.\eeq
Notice the importance of introducing the `infrared' scale $\omega_0$ when studying the gluon
distribution: without such a cutoff, the number of soft gluons would be infinite, as already obvious
from \eqn{Pdeltat}. By varying this cutoff $\omega_0$ one can explore various regimes.
In particular, \eqn{nPoisson} confirms that both the average number of gluons with
$\omega\sim\obr(t)$ and its fluctuations are of order one\footnote{Strictly speaking,
\eqn{nPoisson} refers to the gluons with energies larger than, or equal to, $\omega_0$,
but in fact the averages there are controlled by the softest among these gluons, 
those with $\omega\sim\omega_0$; see the discussion in Sect.~\ref{sec:number}.}.
It also predicts that the average number 
increase quite fast when decreasing $\omega_0$ below $\obr(t)$, while the relative
strength  of the fluctuations decreases: $\sigma_n/\langle n \rangle = 1/\sqrt{\langle n \rangle}
\ll 1$ when $\omega_0\ll\obr(t)$. This last conclusion must be taken with a grain of salt,
since the primary gluons with $\omega\ll\obr(t)$ cannot be distinguished from
the gluons with similar energies which are produced via multiple branching. The latter
are expected to dominate the soft part of the gluon distribution and for them the typical
fluctuations are parametrically large. This will be confirmed 
by the calculations in the next section.

Yet, there are quantities, including observables, which {\em are} sensitive to the primary gluons. One 
such an observable, as we have seen, is the energy loss at large angles; but this probes only
to the relatively hard primary emissions, with $\omega\sim\obr(t)$, so it cannot test the
Poisson distribution \eqref{Poisson} at softer energies. A better candidate in that sense is 
the energy loss $\Delta E$ by the LP, which is by definition controlled by the primary emissions. 
As observed in \cite{Baier:2001yt}, the precise distribution of $\Delta E$ (and not only its average value 
$\langle \Delta E\rangle\sim \abar\omega_c$) is important for understanding the quenching of
the hadronic spectra at large $p_T$. This distribution $\mcal{P}(\Delta E)$ has been computed in
Ref.~\cite{Baier:2001yt} under the assumption that primary emissions are Poissonian. 
Remarkably, the function $\mcal{P}(\Delta E)$ thus obtained coincides with the shape of the LP
peak in the gluon spectrum \eqref{Dexact}, with the identification $\Delta E\equiv E(1-x)$ and for
$1-x\ll 1$. We recall that the spectrum \eqref{Dexact} has been obtained via a more general calculation,
which includes multiple branchings and makes no special assumption about the primary gluons.
In our opinion, this agreement between \eqn{Dexact} with $x\simeq 1$ and the result for $\mcal{P}(\Delta E)$ 
in \cite{Baier:2001yt} provides a strong argument in favor of a Poisson distribution for
the primary emissions. Further evidence in that sense will be presented in the next section.

\section{Event-by-event fluctuations in the gluon distribution}
\label{sec:number}

At several places in the previous discussion, we made remarks about the correlations
in the gluon distribution produced by the branching process. In Sect.~\ref{sec:eqs} we argued that 
gluons should be generally correlated with each other, because they have common ancestors.
Later on, in Sect.~\ref{sec:dis}, we argued that the emissions of primary gluons should be 
independent of each other, because they have no overlap in time. In this section, we would
like to substantiate such previous comments via explicit calculations of the
gluon multiplicity event-by-event and in particular 
demonstrate that they are indeed consistent with each other: gluons can be mutually
correlated, or uncorrelated, depending upon their production
mechanism and upon their energies.

The random variable that we shall consider to that aim is the number $N(t,x_0)$
of gluons with energy fractions larger than some infrared cutoff $x_0\ll 1$ at time $t$,
\begin{align}\label{Nx0def}
N(t,x_0)&\,=\int_{x_0}^1\rmd x
\,\frac{\rmd N}{\rmd x}(t)\,.
\end{align}
Via the correspondence between the low energy cutoff $\omega_0\equiv x_0E$
and the angular opening $\theta_0$ of the jet, namely 
$\sin\theta_0\simeq Q_s/\omega_0$ with $Q_s^2(t)=\hat q t$,
this quantity $N(t,x_0)$ can also be viewed as the event-by-event number
of gluons inside a jet with opening angle $\theta_0$.
This looks similar to the jet energy fraction $X (\tau, x_0)$ introduced in \eqn{Xtau0},
but one should keep in mind that 
these two quantities probe very different aspects
of the gluon distribution inside the jet ($\theta<\theta_0$): the energy fraction
$X (\tau, x_0)$  is controlled by the hardest gluons and notably by the LP, whereas the 
gluon number $N(t,x_0)$ is rather controlled by the softest gluons, with energies
close to the infrared cutoff $\omega_0$. This will become clear after computing the
first two moments of the gluon number distribution, that is,
the mean number
\begin{align}\label{avNdef}
\langle N(t,x_0)\rangle&\,=\int_{x_0}^1\rmd x
\left\langle \frac{\rmd N}{\rmd x}(t)\right\rangle\,=\int_{x_0}^1\frac{\rmd x}{x}\,D(x,\tau)\,,
\end{align}
and its second moment, which also determines the variance (recall \eqn{correlators}),
\begin{align}\label{avN2}
\langle N^2(t,x_0)\rangle&\,=\int_{x_0}^1\rmd x\int_{x_0}^1\rmd x'
\left\langle \frac{\rmd N_{\rm pair}}{\rmd x \,\rmd x'}(t)\right\rangle
+\int_{x_0}^1\rmd x
\left\langle \frac{\rmd N}{\rmd x}(t)\right\rangle
\nonumber\\*[0.2cm]
&\,=\int_{x_0}^1\frac{\rmd x}{x}\int_{x_0}^1\frac{\rmd x'}{x'} D^{(2)}(x,x',\tau)
+\int_{x_0}^1\frac{\rmd x}{x}\,D(x,\tau)\,.
\end{align}
The last equation can be rewritten as $\langle N^2\rangle=\langle N(N-1)\rangle
+\langle N\rangle$, where $\langle N(N-1)\rangle$ is the second factorial moment
(for gluons with $x\ge x_0$, of course).
As before, these quantities will be computed for the simplified kernel $\mcal{K}_0$,
that is, by using Eqs.~\eqref{Dexact} and \eqref{D2exact} for the gluon spectrum
and for the (energy-weighted) gluon pair density, respectively.

As just mentioned, the gluon number distribution is very sensitive to the actual
value of the infrared cutoff $x_0$, hence the importance of properly chosing this value
to match our physical purposes. Clearly, one needs $x_0\ll 1$ in order to describe
radiation. To probe multiple branching at small times $\tau\ll 1$, 
one needs the stronger condition $x_0\lesssim \tau^2$, that is,
$\omega_0\lesssim \obr(t)$. 
For consistency with our approximation scheme, 
this cutoff $\omega_0$ must be harder than the medium scale $T$, 
where the branching dynamics gets modified by elastic
collisions and thermalization \cite{Iancu:2015uja}; hence, we shall choose\footnote{Within
the theory for turbulence, such an intermediate range in energies, which 
is well separated from both the `source' and
the `sink', is known as the `inertial range' \cite{KST,Nazarenko}.}
$T/E\ll x_0\ll 1$. Note that we implicitly 
assume the separation of scales $T\ll  \obr(t)\ll E$, which is indeed
well satisfied in practice.
We furthermore assume that
the medium acts as a `perfect sink' at the low energy end of a
gluon cascade: it efficiently absorbs
the energy carried by the soft branching products with $\omega\lesssim T$ without
modifying the branching dynamics at higher energies $\omega\ge\omega_0$.
The recent study in Ref.~\cite{Iancu:2015uja},
which addressed the interplay between branching and thermalization,
demonstrates that this assumption is rather well satisfied for sufficiently small times
$t\ll\tbr(E)$, or $\tau\ll 1$. 

\subsection{The average gluon number} 
\label{sec:avn}

\begin{figure}[t]
	\centering
	\includegraphics[width=0.6\textwidth]{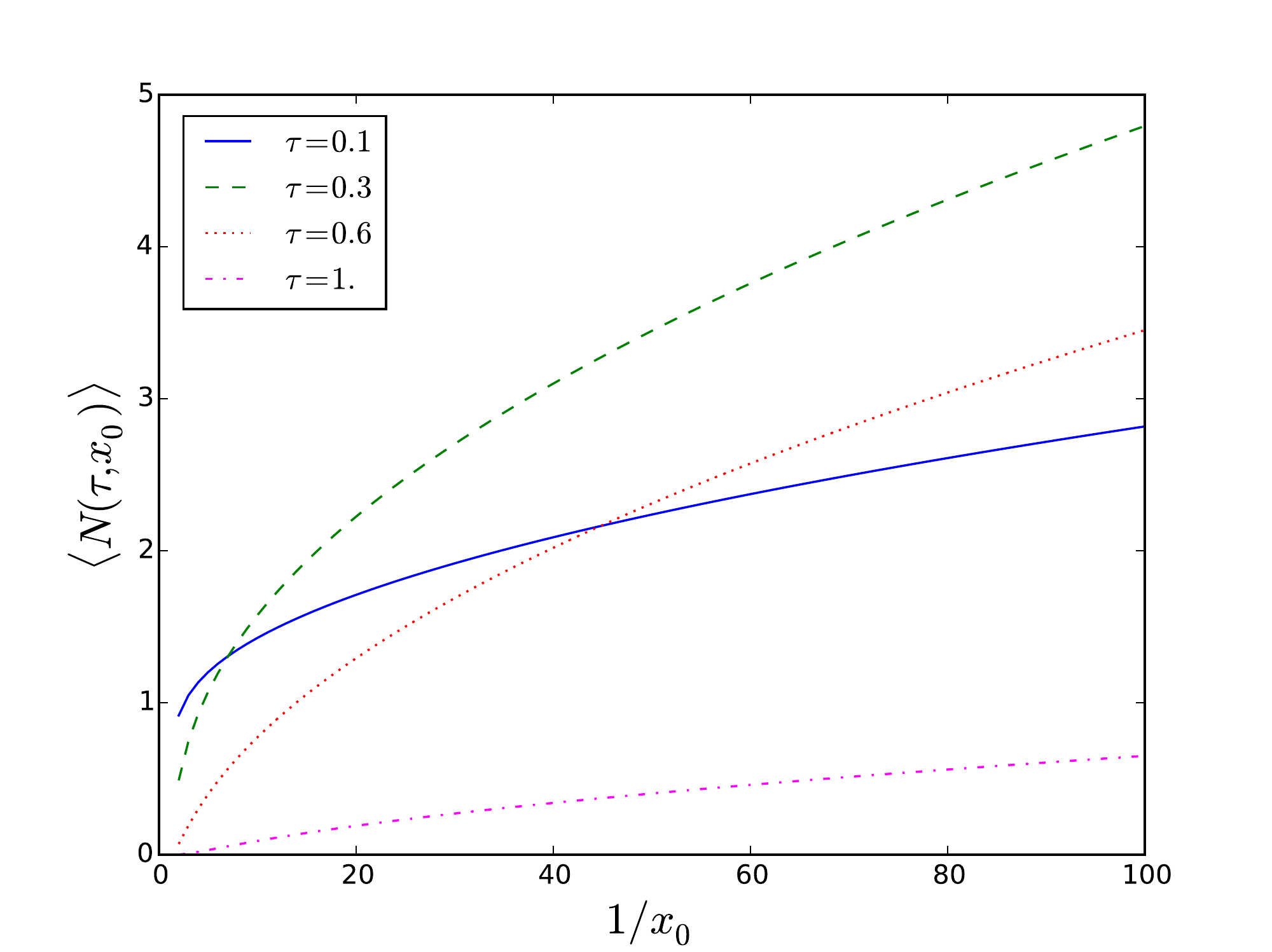}
		\caption{\sl The average number of particles with energy fraction $x\ge x_0$,
		 $\langle N(\tau,x_0)\rangle$, plotted as a function of $1/x_0$ for $x_0\le 0.5$
		 and various values of $\tau$:
		solid (blue): $\tau=0.1$; dashed (green): $\tau=0.3$;
 dotted (red): $\tau=0.6$; dashed--dotted (magenta): $\tau=1$.
		}
		\label{fig:N}
\end{figure}

The calculation of the average gluon number according to \eqn{avNdef} is straightforward and gives 
\begin{align}\label{avn}
\langle N(t,x_0)\rangle&\,=\,
\rme^{-\pi\tau^2}\tau
\int_{u_0}^\infty\frac{\rmd u}{\sqrt{u}}\,\frac{u+1}{u}\,\rme^{-\pi\tau^2 u}
\nonumber\\*[0.2cm]
&=\rme^{-\pi\tau^2}\left[1-\text{erf}\left(\tau\sqrt{\frac{\pi x_0}{1-x_0}}\right)\right]
\left[1-2\pi\tau^2\right]+ {2\tau}\sqrt{\frac{1-x_0}{x_0}}\,\rme^{-\frac{\pi\tau^2}{1-x_0}}
\nonumber\\*[0.2cm]
&\simeq \,
1 +\tau\int_{u_0}^\infty\frac{\rmd u}{u^{3/2}}
\,\simeq \,1 +\frac{2\tau}{\sqrt{x_0}}\,=\,1+2\left[\frac{\obr(t)}{\omega_0}\right]^{1/2}
\,,
\end{align}
where we have changed variables according to $u\equiv x/(1-x)$ [hence,
$u_0= x_0/(1-x_0)$], and the
approximations in the last line hold for $\pi\tau^2\ll 1$ 
and $x_0\ll 1$. 
The above result for $\langle N(t,x_0)\rangle$ is illustrated in
Fig.~\ref{fig:N} for several values of $\tau\le 1$.

The final result at small $\tau$ can be written as 
$\langle N(t,x_0)\rangle\simeq 1+\langle n(t,x_0)\rangle$,
where the 1 refers to the LP, while $\langle n(t,x_0)\rangle={2\tau}/{\sqrt{x_0}}$ is 
the average number of radiated gluons with $x\ge x_0$. 
The above integral yielding $\langle n(t,x_0)\rangle$ is dominated by its lower limit
$u=u_0$, or $x=x_0$; this shows that $\langle n(t,x_0)\rangle$ is essentially the same
as the number of gluons with energy fraction $x_0$.
Hence, by studying the statistics of gluons with an infrared cutoff
$x_0\ll 1$, we actually describe the event-by-event distribution of the particles propagating
at angles $\theta\simeq\theta_0$, with $\theta_0$ as defined after \eqn{Nx0def}.

The above estimate for $\langle n(t,x_0)\rangle$ at small $\tau$ is identical with the respective
prediction, \eqn{nPoisson}, of the Poisson distribution \eqref{Poisson}, which was supposed
to refer to primary gluons only. That is, the average multiplicity of the radiated gluons with 
$x\ge x_0$ is not modified by multiple branching --- it is 
formally the same as produced via radiation by the LP alone. 
This is in line with the previous observation
that the gluon spectrum at $x\ll 1$, \eqn{Dsmallx}, is a fixed point of the branching dynamics.
Of course, the multiple branchings do lead to the copious production of soft gluons with 
$x\lesssim \tau^2$, but these gluons decay as fast as they are produced 
and leave behind only {\em very} soft gluons, with energies $\omega\lesssim T$,
which disappear in the plasma (formally, they accumulate into a
condensate at $x=0$). Such gluons are not counted in our estimate for $\langle N(t,x_0)\rangle$, which is valid only for $x_0>T/E$.

Another interesting consequence of the final result in \eqn{avn} is that
the number $\langle n(t,\omega_0)\rangle$ of soft radiated gluons is independent
of the energy $E$ of the LP (so long as $\omega_0\ll E$ and for sufficiently small times $t\ll \tbr(E)$). 
This number becomes large, $\langle n(t,\omega_0)\rangle > 1$, 
when the cutoff is sufficiently soft, $\omega_0\lesssim\obr(t)$.
We thus conclude that a high-energy jet with $E\gg\obr(L)$ 
produces a large number of soft 
gluons which propagate at large angles;
this number is roughly independent of $E$ and grows linearly with the distance
$L$ travelled by the jet through the medium.

Note finally that $\langle n(t,\omega_0)\rangle$ is rapidly increasing when decreasing
the energy cutoff $\omega_0$, or, equivalently, when increasing the jet opening angle 
$\theta_0 \propto 1/\omega_0$ (see Fig.~\ref{fig:N}).  This should be contrasted to
the respective behavior of the jet energy, which is almost independent of
$\omega_0$ (or $\theta_0$) when $\omega_0$ is small enough  (recall Fig.~\ref{fig:X}). 
This difference is easy to understand: when decreasing
$\omega_0$ (or increasing $\theta_0$), one includes new gluons within the jet, 
whose number is increasing as $1/\sqrt{\omega_0}$ (or equivalently like $\sqrt{\theta_0}$), 
but which carry only little energy $\propto \sqrt{\omega_0}$.

\subsection{The gluon number fluctuations}
\label{sec:NN}

We now turn to the gluon number correlations, as measured by the second factorial moment 
\begin{align}\label{avNN}
\langle N(N-1)\rangle(\tau,x_0)\,=\int_{x_0}^1\frac{\rmd x}{x}\int_{x_0}^{1-x}\frac{\rmd x'}{x'} D^{(2)}(x,x',\tau)
\,,
\end{align}
with $D^{(2)}(x,x',\tau)$ given by \eqn{D2exact}.
This calculation turns out to be more complicated and we have not succeeded in obtaining an
exact result in closed form. But for the most interesting physical regimes, we have obtained
accurate results with a transparent physical interpretation. In fact it is easy to anticipate
these results via simple arguments that we now present. As before, we assume $x_0\ll 1$
and focus on relatively small times $\tau\ll 1$, when the LP still exists.

The factorial moment 
$\langle N(N-1)\rangle(\tau,x_0)$ counts twice the number of gluon pairs in which both gluons 
have energy fractions larger than, or equal to, $x_0$.
Clearly, there are two types of such pairs:  (A) those made with the LP plus a radiated 
gluon, and (B) those in which both gluons are radiation products. It is furthermore clear that the average 
number of pairs of types A is the same as the average number of radiated gluons with $x\ge x_0$, 
that is, $\langle N(N-1)\rangle_{\rm A}= 2\langle n(\tau,x_0)\rangle\simeq {4\tau}/{\sqrt{x_0}}$, cf. \eqn{avn}. 
Concerning the pairs of type B, we expect their number
to be dominated by those pairs in which both gluons are soft, meaning that the integrals
in \eqn{avNN} are controlled by $x,\,x'\ll 1$. In this  regime, 
one can use the simplified version of the pair density in \eqn{D2small}, to deduce
\begin{align}\label{NNx0}
\langle N(N-1)\rangle_{\rm B}&\,\simeq\,\frac{3}{2}\,\tau^2
\int_{x_0}^1\frac{\rmd x}{x^{3/2}}\int_{x_0}^1\frac{\rmd x'}{(x^\prime)^{3/2}} 
\,\simeq\,\frac{6\tau^2}{x_0}
\,,
\end{align}
where the integrals are indeed controlled by the lower limit (hence, the upper limit has
been arbitrarily set to 1).  Adding the previous results, one finds
\begin{align}\label{NNav}
\langle N(N-1)\rangle(\tau,x_0)\,\simeq\,\frac{4\tau}{\sqrt{x_0}}\,+\,\frac{6\tau^2}{x_0}\,.
\end{align}
This turns out to be indeed the right result, at least in the two limiting situations where
we have been able to approximately compute the integrals in \eqn{avNN}, namely 
 \texttt{(i)} the multiple branching regime at $x_0\ll\tau^2\ll 1$ and  \texttt{(ii)} 
 the perturbative regime at $\tau^2\ll x_0\ll 1$ (see App.~\ref{app:N2} for details).
The transition region around $x_0=\tau^2$ turns out to be more difficult to study, but
the fact that we obtain identical results
when approaching this region from both sides makes us confident that the
transition is quite smooth and reasonably well described by  \eqn{NNav}.

The r.h.s. of \eqn{NNav} coincides with the respective prediction
of the perturbative expansion in the number of branchings to second order: 
the first term, linear in $\tau$, is the same
as the result of a single emission by the LP, while the second term, quadratic in $\tau$, 
would be also obtained by summing the 3 processes which involve exactly 2 
branchings, cf. Fig.~\ref{fig:2g}. 
We thus conclude that the 2nd
factorial moment is not modified by multiple branchings, similarly to the mean
multiplicity.
Once again, this feature is a non-trivial consequence of the wave turbulence, 
i.e. of the fact that the 
energy flux generated via democratic branchings is independent of $x$.

Using  \eqn{NNav} together with \eqn{avn}, one can immediately compute
the variance in the number of gluons with $x\ge x_0$~:
\begin{align}\label{Nvar}
\sigma^2_N(\tau,x_0)&\,\equiv\,\langle N(N-1)\rangle + \langle N\rangle - \langle N\rangle^2
\,\simeq\,\frac{2\tau}{\sqrt{x_0}}\,+\,\frac{2\tau^2}{x_0}
\,.
\end{align}

Note that, when rewritten in terms of the dimensionfull quantities $t$  and
$\omega_0$, both the second moment \eqref{NNav} and the variance \eqref{Nvar}
scale like functions of $\obr(t)/\omega_0\propto t^2/\omega_0$ 
and are independent of the energy $E$ of the LP.
A similar behavior has been observed for the average multiplicity \eqref{avn}.

The physical consequences of these results will be now separately discussed for the
two limiting regimes aforementioned.


\subsubsection{$x_0\ll\tau^2\ll 1$ :  Koba-Nielsen-Olesen scaling}

In this soft regime, the typical gluon numbers are large and their fluctuations
are relatively large as well, as we shall shortly see. Specifically,
the calculations in App.~\ref{app:N2} yield (cf. \eqn{ap:NNsmall})
\begin{align}\label{NNsmall}
\langle N(N-1)\rangle(\tau,x_0)\,=\,\frac{6\tau^2}{x_0}\,+\,\frac{4\tau}{\sqrt{x_0}}\,+\,
\mathcal{O}\left(1,\frac{\tau^4}{x_0}\right).
\end{align}
Besides confirming the previous estimate in \eqn{NNav}, this result
also specifies the respective error\footnote{The calculations in
App.~\ref{app:N2} also confirm the physical origin of the 2 terms in the r.h.s.
of \eqn{NNsmall}: the quadratic term is controlled by
very soft gluons, with $x\sim x'\sim x_0$;
the linear term arises by integrating over $x\sim x_0$ and 
$1-x'\simeq\pi\tau^2$, or vice-versa; that is, this term measures the
pairs made with one soft gluon and the LP.}.
Clearly, in this regime, the term quadratic in $\tau$ in the r.h.s. of
\eqn{NNsmall} dominates over the linear one.
By keeping only this dominant term, together
with the corresponding approximation $\langle N\rangle\simeq {2\tau}/{\sqrt{x_0}}
\gg 1$ for the average gluon number, one can write
\beq\label{NNcum}
\langle N(N-1)\rangle(\tau,x_0)
\,=\,\frac{3}{2}\,\langle N(\tau,x_0)\rangle^2\,+\,\order{\frac{\tau}{\sqrt{x_0}}}\,.
\eeq
This shows that 
the second factorial cumulant $\langle N(N-1)\rangle  - \langle N\rangle^2$,
which we recall is a measure of the correlations, is parametrically as large as
the `disconnected' 2-point function $\langle N\rangle^2$ (and much larger 
than the mean number $\langle N\rangle$).  This means that
correlations are large. Recalling the physical origin of the term quadratic
in $\tau$, cf.  \eqn{NNx0}, it is clear that the net correlation, namely
\beq\label{Ncor}
\langle N(N-1)\rangle  - \langle N\rangle^2\,\simeq\,
\frac{1}{2}\,\langle N(\tau,x_0)\rangle^2\,\simeq\,
\frac{2\tau^2}{x_0}\,,
\eeq
comes from processes in which the two measured gluons belong to
a same mini-jet, i.e. they have a {\em soft} common ancestor, cf. Fig.~\ref{fig:correl} 
left. The contributions to  $\langle N(N-1)\rangle$ coming from two different mini-jets,
cf. Fig.~\ref{fig:correl} right, cancel against the disconnected piece 
$\langle N\rangle^2$. Soft gluons which belong to different mini-jets are 
uncorrelated with each other, because so are the primary gluons which have initiated those mini-jets in the first place (see Sect.~\ref{sec:Poisson} below).  

To the accuracy of interest, the variance can be read from either \eqn{Ncor}, or
\eqn{Nvar}; one thus finds $\sigma^2_N(\tau,x_0)\simeq {2\tau^2}/{x_0}$.
This is large, $\sigma^2_N \gg \langle N\rangle \gg 1$,
showing that the gluon distribution at small $x$ is {\em overdispersed}
(as compared to the Poisson distribution, which would predict $\sigma^2_N = \langle N\rangle$).

The above results can be summarized as
\beq
\label{dispersionN}
\sigma_{N}(t,\omega_0)\,\simeq\,\frac{1}{\sqrt{2}}\,\langle N(t,\omega_0) \rangle
\,\simeq\,\left[\frac{2\obr(t)}{\omega_0}\right]^{1/2}\qquad\mbox{when\quad 
$t\ll \tbr(E)$ and $\omega_0\ll\obr(t)$}
\,,\eeq
where we have restored the physical units to emphasize that, to the accuracy of interest,
the result is independent of $E$. \eqn{dispersionN} is similar to \eqn{dispersion},
in that it shows that the typical fluctuations in the gluon number distribution at small $x$
are as large as the respective average value.

We expect factorization properties similar to \eqn{NNcum}
also for the higher moments:
\beq\label{Cp}
\langle N(N-1)(N-2)\cdots (N-p+1)\rangle\,\simeq \,\langle N^p\rangle
\,\simeq \,C_p \langle N\rangle^p\,,
\eeq
where the factors $C_p$ (known as `reduced moments') are pure numbers.
Indeed, as shown by the example of the 2-point function \eqref{D2small},
it is natural to expect the factorization of the $p$-point correlation 
function $D^{(p)}(x_1,x_2, ...,x_p)$ for small values of its arguments, $x_i\ll 1$:
the soft gluons evolve independently from each other after the splitting of
their last common ancestor. Notwithstanding, the correlations are strong, 
because the information about the LCA cannot be lost: it determines the topology
of the cascade.

A probability distribution for which the (factorial) moments of order $p$ scale like
the $p$-th power of the mean number is said to obey {\em  KNO scaling} \cite{Koba:1972ng}.
This is tantamount to saying that the product $\langle N\rangle \mcal{P}_N$ between the mean number
and the respective probability distribution $\mcal{P}_N$ depends upon
the random variable $N$ and the various parameters of the distribution (in our case $\tau$ and $x_0$) 
only via the ratio $N/\langle N\rangle$. This has interesting consequences for the phenomenology: the ratio $\langle N^2\rangle/\langle N\rangle^2$ is
predicted to be independent of $\tau$ and $x_0$ 
(and similarly for the higher reduced moments $C_p$). This could be explicitly
checked in the data --- say, in the asymmetric di-jet events in $AA$ collisions ---, 
by measuring the event-by-event distribution of soft gluons propagating at large angles.

One additional reason why the KNO scaling looks 
appealing in the present context is because a similar scaling (albeit with a different
value for the second reduced moment $C_2$; see below) has been observed
to hold for jets evolving in the vacuum \cite{Dokshitzer:1991wu}. In that case, the
branchings are triggered by the virtuality of the leading particle, so
the branching probability is different --- it is given by the DGLAP splitting functions.
This difference has indeed consequences for the details of the distribution, like
the values of the reduced moments $C_p$. For jets evolving in the vacuum, one 
finds $C_2=4/3$ in the leading double-logarithmic approximation
\cite{Dokshitzer:1991wu}. That is, the vacuum analog of 
our \eqn{NNcum} reads\footnote{An
elementary probability distribution which is known to lead to KNO scaling in the regime
of large numbers is the {\em negative binomial distribution} (NBD), whose definition
involves a  parameter $r$. The relations \eqref{NNcum} and \eqref{NNvac} are
consistent with the NBDs with  $r=2$ and $r=3$, respectively. Yet, we see no physical
reason why the NBD with $r=2$ should truly apply to the gluon distribution produced
via medium-induced branchings. Indeed, for a jet decaying in the vacuum, 
for which all the factorial moments can be explicitly computed, the NBD with $r=3$
fails to describe these moments for large values of $p$ 
--- that is, it does not reproduces the respective values of $C_p$
for $p\ge 3$ \cite{Dokshitzer:1991wu}.}
\beq\label{NNvac}
\langle N(N-1)\rangle\,=\,\frac{4}{3}\,\langle N\rangle^2\quad
\mbox{(jets in the vacuum)}.
\eeq
This vacuum distribution looks already quite broad (say, as compared to a Poisson distribution,
for which $\langle N(N-1)\rangle= \langle N\rangle^2$),
but the one generated via medium-induced branchings, for which $C_2=3/2$, is {\em even broader}. 
So, the fluctuations in the gluon distribution produced by a jet propagating through a dense 
medium should look even larger (in appropriate units, cf. \eqn{Cp}) 
than for a jet which propagates in the vacuum. 

\subsubsection{$\tau^2\ll x_0\ll 1$ : primary gluons and the Poisson distribution}
\label{sec:Poisson}

In this regime, gluon branchings have a low probability, hence
the average number of radiated gluons is small,
$\langle n(t,x_0)\rangle\simeq {2\tau}/{\sqrt{x_0}}\ll 1$, and fully controlled
by direct emissions by the LP: these are {\em primary} gluons.
Concerning the second moment,
the calculations in App.~\ref{app:N2} yield (cf. \eqn{ap:NNlarge})
\begin{align}\label{NNlarge}
\langle N(N-1)\rangle(\tau,x_0)\,=\,\frac{4\tau}{\sqrt{x_0}}\,+\,\frac{6\tau^2}{x_0}\,+\,
\mathcal{O}\left(\frac{\tau^3}{x_0^{3/2}}\right),
\end{align}
where the linear (quadratic) term represents the leading (subleading) contribution.
The perturbative expansion in the number of branchings is now meaningful 
and the two terms in the r.h.s. of \eqn{NNlarge} 
can be {\em physically} (and not only formally) recognized as the results of 
a single emission (for the term linear in $\tau$)
and, respectively, of a double emission (for the quadratic term).

The variance too is dominated by the linear term in the r.h.s. of \eqn{Nvar} and hence
it is equal to the average number $\langle n(t,x_0)\rangle\simeq {2\tau}/{\sqrt{x_0}}$ of 
primary gluons.
This is consistent with our conclusion in Sect.~\ref{sec:dis} that the primary 
emissions with $x\ll 1$ should obey a Poisson distribution, 
cf. Eqs.~\eqref{Poisson}--\eqref{nPoisson}.
This argument can be made more precise, as we now explain.

To that aim, notice that the quadratic term in the r.h.s. of \eqn{NNlarge} receives
2/3 of his strength from processes involving the emission of 2 primary gluons,
cf. Fig.~\ref{fig:2g}. Hence, after subtracting 1/3 of this term (which is associated with
the production of 2 secondary gluons, as shown by the first process in Fig.~\ref{fig:2g}),
the remaining contribution, that is,
 \beq \langle N(N-1)\rangle\Big |_{\rm primary} 
 \simeq \,\frac{4\tau}{\sqrt{x_0}}\,+\,\frac{4\tau^2}{x_0}\,,\eeq
involves only primary emissions. For such emissions,
it makes sense to decompose the random variable $N$ as $N=1+n$
(the LP plus the number $n$ of primary gluons).
We then easily deduce
  \begin{align}\label{nnlarge}
\langle n(n-1)\rangle\,=\,\langle N(N-1)\rangle\Big |_{\rm primary} - 2\langle n\rangle
\,\simeq\,\frac{4\tau^2}{x_0}\,\simeq\,\langle n\rangle^2
\,,
\end{align}
or, equivalently, $\sigma^2_n=\langle n\rangle$, which is the hallmark of the Poisson
distribution, cf. \eqn{nPoisson}.

\comment{
Note finally that the above mathematical argument can be formally repeated for the
gluon distribution at small $x_0\lesssim\tau^2$, as produced via multiple branchings:
after subtracting 1/3 from the quadratic term in \eqn{NNsmall}, the remaining contribution
to $\langle N(N-1)\rangle$ is consistent with a Poisson distribution for the
soft gluons. This reflects the fact
that even soft gluons which belong to different mini-jets are independent from each other,
because so are the primary gluons which have initiated those mini-jets in the first place. 
Vice-versa, the net correlation among the soft gluons (as represented
by 1/3 of the quadratic term in \eqn{NNsmall}) comes from processes where the
soft gluons belong to a same mini-jet (cf. Fig.~\ref{fig:correl} left).
}
 
 \section{Summary and conclusions} 
 \label{sec:conclusion}

In this paper, we have for the first time investigated the effects of event-by-event fluctuations
in the gluon distribution produced via medium-induced branching by an energetic jet propagating
through a dense QCD plasma. We have identified a characteristic pattern for the 
medium-induced radiation, which can be observed on an event-by-event basis, 
albeit the fluctuations from one event to another are predicted to be large. This pattern is
characterized by the production of many soft gluons --- most of them, as soft as the medium
scale $T$ --- which propagate at large angles w.r.t. the jet axis and collectively carry a 
large amount of energy. 

The overall picture depends upon the ratio between the energy $E$
of the leading particle and the medium scale $\obr(L)=\abar^2\hat q L^2$ (the characteristic
energy for multiple branching), or, equivalently, between the branching time 
$\tbr(E)=(1/\abar)\sqrt{E/\hat q}$ for the LP and the distance $L$ travelled by the jet through 
the medium. For not so high energies, such that $E\lesssim \obr(L)$ (or $\tbr(E)\lesssim L$),
the LP disappears via democratic branching and its whole energy is ultimately transmitted
to soft gluons propagating at large angles. In the high-energy regime at $E\gg\obr(L)$
(or $\tbr(E)\gg L$) --- the most interesting regime for the phenomenology of di-jet asymmetry
at the LHC --- the LP survives in the final state and the energy transferred at large angles  
in a typical event is independent of $E$ and of the order of the medium scale $\obr(L)$. 
Moreover, the event-by-event 
fluctuations in the energy loss at large angles are of order $\obr(L)$ as well --- that is, they are
as large as the respective average value.

The physical interpretation of this result is quite interesting: the energy gets transmitted to
softer and softer quanta via successive quasi-democratic branchings. The energy flux
generated by such branchings is independent of $\omega$ (at least, within the inertial
range at $T\ll\omega \ll E$). Hence the energy loss
at large angles is controlled by the hardest partons which can undergo democratic 
branching during a time $\sim L$ --- the primary gluons with energies $\omega\sim\obr(L)$.
Such gluons are emitted by the LP with a probability of order one and their 
emissions are quasi-independent of each other. Hence, both the average number of
such gluons and its typical fluctuations are of $\order{1}$. In turn, this implies that both
the average energy loss at large angles and the respective dispersion should be 
of order $\obr(t)$, which is what we found (cf. \eqn{dispersion}).

Another remarkable feature of the gluon distribution $x(\rmd N/\rmd x)$ produced  
by an energetic jet with $E\gg\obr(L)$
is the fact that, for the soft gluons with $x=\omega/E\ll 1$, this depends only upon
the dimensionless ratio  ${\tau}/\sqrt{x}=[\obr(L)/\omega]^{1/2}$. In particular, it
is independent of $E$. This property too is a consequence of wave turbulence, i.e. of the fact 
that the gluon distribution at small $x$ is not modified by multiple branching. For instance, 
the gluon spectrum $D(x)$ with $x\ll 1$ has the scaling form visible in \eqn{Dsmallx}, 
while the 2-point function $D^{(2)}(x,x')$ with $x,\,x'\ll 1$ factorizes as shown in
\eqn{D2small}. A similar factorization property is expected for the higher-point correlations.
It reflects the fact that soft gluons evolve independently from each other 
after the splitting of their last common ancestor.

These special properties of the gluon distribution at small $x$ have interesting
consequences for the event-by-event number $N(\omega_0)$ of gluons
with energies $\omega \ge \omega_0$. First,  the multiplicity distribution too is
independent of $E$ and solely a function of $\obr(L)/\omega_0$.
Second, in the non-perturbative regime at $\omega_0 \ll \obr(L)$, where the
multiplicity is large, the higher moments $\langle N^p(\omega_0)\rangle$ are predicted to
obey KNO scaling. For instance, the ratio $C_2\equiv \langle N^2\rangle/
\langle N\rangle^2$ should be a pure number, independent of all the physical parameters
of the problem (like $E$, $L$, $\hat q$, or the infrared cutoff $\omega_0$).
This strong prediction should be taken with a grain of salt, as it assumes that the only
source of fluctuations is the stochasticity of the branching process.
In the actual collisions though, there are also other stochastic aspects,
which in particular imply that the distance $L$ travelled by the jet trough the medium
is itself a random variable. Hence equations like \eqref{NNcum} or \eqref{VAR}
should be more properly read as
\beq
\frac{\langle N^2(\omega_0)\rangle}{\langle N(\omega_0)\rangle^2}\,\simeq\,\frac{3}{2}\,
\frac{\langle L^2\rangle}{\langle L\rangle^2}\,,\qquad
\frac{\langle\mcal{E}^2\rangle}{\langle\mcal{E}\rangle^2}
\,\simeq\,\frac{4}{3}\,
\frac{\langle L^4\rangle}{\langle L^2\rangle^2}
\,,
\eeq
where the various averages also include the event-by-event distribution of $L$.
Such relations are both interesting and non-trivial: not only they are still independent
upon physical parameters like $E$, $\hat q$, or $\omega_0$, but they also show that,
by combining event-by-event measurements of the soft hadron distribution at large angles
with the results of the present theoretical analysis, one can better constrain the distribution of $L$
in the experiments.

\comment{Our results demonstrate that the medium-induced gluon branching leads to a
characteristic pattern for the radiation, which is very different, and also geometrically
separated --- in the sense that the medium-induced radiation propagates at large angles ---
from the vacuum-like radiation associated with the virtualities of the jet constituents.
This geometric separation is important, as it implies that in an actual experiment,
where both sources of radiation coexist with each other, their effects can still
be disentangled, by looking at the (energy and hadron number) distribution at large angles.
}

We conclude with a few numerical estimates, which illustrate the possible implications of our 
analysis for the phenomenology.  We consider a leading particle with $E=100$\,GeV
which propagates through the medium along a distance $L$ for which we shall consider
3 representative values: $L=\{2;\, 4;\, 6\}\,$fm. As before, we chose $\abar=0.3$
and $\hat q= 1$\,GeV$^2/$fm. With these choices, one finds $\tbr(E)\simeq 15\,$fm,
hence the `high-energy' (or `small medium size') condition $\tbr(E)\gg L$ is
reasonably well satisfied for all the interesting values of $L$. Corresponding to the 3 values
for $L$ aforementioned, we find: $\obr(L)\simeq \{2;\, 8;\, 18\}\,$GeV. Using \eqn{dispersion},
the average energy loss at large angles and the associated dispersion are roughly estimated as   
$\langle\mcal{E}(L)\rangle\simeq \{6;\, 25;\, 57\}\,$GeV
and, respectively, $\sigma_{\mcal{E}}(L)\simeq \{3;\, 15;\, 33\}\,$GeV. This shows that,
e.g. for $L=4$\,fm, the expected range for the variation of the event-by-event energy loss
is $10 < \mcal{E} < 40\,$GeV.

Consider similarly the statistics of the number of gluons with energy $\omega\ge \omega_0$,
with $\omega_0$ chosen of the order of the medium temperature $T$, say  $\omega_0=0.5$\,GeV.
The mean number of radiated gluons for the 3 choices of the medium size 
$L$ is then estimated as $\langle n(L,\omega_0)\rangle=
2\left[{\obr(L)}/{\omega_0}\right]^{1/2}\simeq \{4;\, 8;\, 12\}$. Note that
this number grows linearly with $L$.
As explained after \eqn{avn}, most of these gluons are soft, say with energies below 
2\,GeV, hence they propagate at large angles, outside the reconstructed jet in the final state.
The variance is estimated according to \eqn{Nvar} as
$\sigma^2_N(L,\omega_0)\simeq \{12;\, 40;\, 84\}$. Hence, for  $L=4$\,fm
we expect $n(\omega_0)$ to take values within the range $2 < n(\omega_0) < 14$.
In an actual experiment, these numbers should further increase due to the gluon
fragmentation into hadrons.

Quite remarkably, the above numbers agree quite well with 
the phenomenology of di-jet asymmetry at the LHC
 \cite{Aad:2010bu,Chatrchyan:2011sx,Chatrchyan:2012nia,Aad:2012vca,Chatrchyan:2013kwa,Chatrchyan:2014ava,Aad:2014wha,Khachatryan:2015lha,Khachatryan:2016erx}. 
 Such an agreement
should be taken with a grain of salt, since our present description is rather idealized: 
we have ignored several important dynamical ingredients, 
like the elastic collisions off the medium constituents, 
the longitudinal expansion of the medium, the virtualities of the partons in the jet, 
the gluon fragmentation into hadrons, and the additional sources of fluctuations, 
besides the randomness of the branching process.
Clearly, all such effects are needed to achieve a complete and
accurate description of the data. Notwithstanding, we expect our simplified picture to
already capture
the main dynamics responsible for energy transfer at large angles, where the vacuum-like
radiation contributes only little. Therefore, the fact that we already observe a reasonable 
agreement with the data is encouraging and calls for more
elaborated studies, which should complete the present picture of multiple branching 
with the additional dynamical ingredients that are still missing.
 
\section*{Acknowledgments}
\vspace*{-0.3cm}
We would like to thank Al Mueller for inspiring discussions and
for a careful reading of the manuscript. One would us (E.I.) acknowledges 
related discussions with Jean-Paul Blaizot and Yacine Mehtar-Tani 
during the early stages of this work.
This work is supported by the European Research Council 
under the Advanced Investigator Grant ERC-AD-267258 and by the 
Agence Nationale de la Recherche project \# 11-BS04-015-01. 

\appendix
\section{Master equations for the Markovian branching process}
\label{app:Markov}

In this Appendix, we shall construct the transport equations \eqref{eqD} and
\eqref{eqD2} obeyed by the energy density $D(x,t)$ and by the gluon pair density
$D^{(2)}(x,x',t)$, respectively. To that aim, we start with the master equations
obeyed by the probability densities $ \mcal{P}_n(x_1,x_2,\cdots,x_n| t)$ introduced
in Sect.~\ref{sec:eqs}. These equations can be easily established via elementary probabilistic
considerations. We recall that $ \mcal{P}_n(x_1,x_2,\cdots,x_n| t)$ is the semi-inclusive
probability density for having a state with $n$ gluons with energy fractions $x_i\ge\epsilon$
($i=1,\dots,n$), together with an arbitrary number of softer gluons with $x<\epsilon$,
which are not measured. The dependence upon the infrared cutoff $\epsilon$ will be
generally omitted at intermediate steps since the final equations for the correlations 
admit a well-defined limit $\epsilon\to 0$.

In one evolution step, $t\to t+\rmd t$, this probability $ \mcal{P}_n$ can increase due to the
splitting of a gluon from a preexisting state with $n-1$ gluons, but it can also decrease via the
decay of one of the $n$ gluons explicitly included in $ \mcal{P}_n$. The splitting rate can be read
off \eqn{Pdef}, conveniently rewritten as 
\beq\label{Ibr}
\frac{\rmd^2 \mcal{I}_{\rm br}}{\rmd z\,\rmd \tau}\,=\,\frac{\mcal{K}(z)}{2\sqrt{x}}\,\equiv\,
\mcal{K}(z,x)\,,\qquad\mbox{with}\ \ \tau\equiv \frac{t}{\tbr(E)}\,,
\eeq
where we recall that $x=\omega/E$ is the energy fraction of the parent gluon and
$z$ is the splitting fraction; that is, the 2 daughter gluons have energies $z\omega$
and $(1-z)\omega$, respectively. 
These considerations motivate the following evolution law for
$\mcal{P}_n$, involving a {\em loss term} and a {\em gain term}~:
 \begin{align}\label{master}
 \frac{\del \mcal{P}_n}{\del\tau}\,=\,&-\left[\sum_{i=1}^n  \int\rmd z \, \mcal{K}(z,x_i)\right]
  \mcal{P}_n(x_1,x_2,\cdots,x_n |\tau)\nn
  &+  \int\rmd z \int\rmd x'\,\mcal{K}(z,x')\nn
  &\quad \times \sum_{i=1}^{n-1}\mcal{P}_{n-1}(x_1 \dots x_{i-1}, x',
  x_{i+1}\dots x_{n-1}|\tau)\,\delta(x_i-zx')\,\delta(x_n-(1-z)x')\,.\end{align}
Recalling equations \eqref{Pdef} and \eqref{Ibr}, it is quite clear that
the integral over $z$ in the negative, `loss', term has endpoint singularities at $z=0$
and $z=1$. These are regulated by the conditions that both daughter gluons be harder
than the infrared cutoff: $zx>\epsilon$ and $(1-z)x>\epsilon$.

Given the master equation \eqref{master} and the rule \eqref{aveO} for computing
the expectation value $\langle \mcal{O}(\tau)\rangle$ of a generic observable, it is straightforward
(albeit a bit tedious) to deduce an evolution equation for the latter. This is conveniently written as
 \begin{align}\label{Oevol}
 \frac{\del\langle \mcal{O}(\tau)\rangle}{\del\tau} \,=\,
 &\sum_{n=1}^{\infty}\int \prod_{i=1}^n \rmd x_i\,\mcal{P}_n(\{x\}|\tau)
 \ \sum_{i=1}^{n} \int\rmd z\,\mcal{K}(z,x_i)\,
 \Delta \mcal{O}_{n; \,i}(\{x\},z)
 \,,\end{align}
with the notations $\{x\}=(x_1,x_2, \dots x_n)$ and
\beq\label{DeltaO}
 \Delta \mcal{O}_{n; \,i}(\{x\},z)\,\equiv\,
 \mcal{O}_{n+1}(x_1 \dots x_{i-1}, zx_i,x_{i+1}\dots x_{n},(1-z)x_i)-
 \mcal{O}_{n}(\{x\})\,.
 \eeq
The subscript $i$ in  \eqn{DeltaO} refers to the gluon $x_i$ which splits in the evolution
step under consideration.
Note that, in general, \eqref{Oevol}  is not a closed equation, that is, the r.h.s. cannot be written
as a linear operator acting on  $\langle \mcal{O}(\tau)\rangle$. 

As a first example, consider
the gluon spectrum $D(x,\tau)$, as defined in \eqn{correlators}.
 In this case, $D_n(\{x\};x)=x\sum_i^n\delta(x_i-x)$ and
\beq\label{DeltaD}
 \Delta D_{n; \,i}(\{x\},z; x)\,=\,x\left[\delta(zx_i-x) + \delta((1-z)x_i-x) - \delta(x_i-x)\right] \,.\eeq
Note that, in this case, the r.h.s. of the above equation is ``local in $i$'' --- that is, it depends
upon the $n$-gluon configuration $\{x\}=(x_1,x_2, \dots x_n)$ only via the energy fraction $x_i$
of the gluon which has split in this particular evolution step. Accordingly, after inserting
\eqn{DeltaD} into the general equation \eqref{Oevol}, one finds a {\em closed} equation
for the gluon spectrum, namely
 \begin{align}\label{eqD0}
   \frac{\del D(x,\tau)}{\del\tau}\
  &=\int \rmd z \,\bigg\{{\cal K}\Big(z, \frac{x}{z}\Big)
  D\Big(\frac{x}{z}, \tau\Big)+
  {\cal K}\Big(z, \frac{x}{1-z}\Big)
  D\Big(\frac{x}{1-z},\tau\Big)-
  {\cal K}(z, x)
  D\big({x},\tau\big)\bigg\}.
  \end{align}
In the first `gain' term, the integral over $z$ is restricted to $z\ge x$ [by the
support of  $D(x/z,\tau)$] and has an endpoint singularity at $z=1$.
In the second `gain' term, we similarly have $1-z\ge x$ and a singularity at $z=0$.
Finally, the negative, `loss', term has two endpoint singularities, at $z=0$ and $z=1$.
Each of these terms could be individually regulated by the infrared cutoff $\epsilon$,
but this is actually not needed: the singularities mutually cancel in the sum
of the three terms, so the complete equation admits a well-defined limit 
$\epsilon\to 0$, as anticipated.
Using the symmetry property $ {\cal K}(z, x)= {\cal K}(1-z, x)$, 
it is clear that the two `gain' terms give identical contributions.
As for the `loss' terms, this can be rewritten in a more convenient form by using
$\mcal{K}(z,x)=\mcal{K}(z)/2\sqrt{x}$ together with the identity 
 \beq\label{endpointssing}
  \int_0^1\rmd z \,{\cal K}(z) \,=\,2 \int_0^{1} \rmd z \,z\,{\cal K}(z) 
  \,=\,2 \int_0^{1} \rmd z \,(1-z)\,{\cal K}(z) 
 \,.\eeq
Putting things together, we recognize \eqn{eqD}, as expected. 

The evolution of the gluon pair density $D^{(2)}(x,x',\tau)$ could be similarly treated,
but the corresponding argument is considerably more tedious. An alternative formulation,
which is more convenient in practice, relies on the following {\em generating functional}
 \beq\label{Zdef}
Z_\tau[u(x)]\,\equiv\,\sum_{n=1}^{\infty}\int \prod_{i=1}^n \rmd x_i\,\mcal{P}_n(\{x\}|\tau)\,
 u(x_1)u(x_2)\dots u(x_n)\,,\eeq
with $u(x)$ an arbitrary `source' function with support at $0\le x\le 1$. Probability
conservation requires $Z_\tau[u=1]=1$.  It is easily checked
that the correlation functions of the gluon density can be obtained as
functional derivatives of $Z_\tau[u]$ evaluated at $u(x)=1$. For instance
 \beq\label{DZ}
D(x,\tau)\,=\,x\,\frac{\delta Z_\tau[u]}{\delta u(x)}\bigg|_{u=1}\,,\qquad
D^{(2)}(x,x',\tau)\,=\,x x'\,\frac{\delta Z_\tau[u]}{\delta u(x)\delta u(x')}\bigg|_{u=1}\,.
 \eeq
Clearly, the generating functional can be viewed as a special kind of `observable',
hence it obeys an evolution equation with the structure shown in \eqn{Oevol}.
Using the latter together with simple manipulations, one finds (see also \cite{Blaizot:2013vha}
for a more general equation of this type, which includes the effects of transverse diffusion)
  \beq\label{eqZ}
  \frac{\del Z_\tau[u]}{\del\tau}\,=\,
  \int\rmd z \int\rmd x\,\mcal{K}(z,x)\left[u(zx) u((1-z)x)- u(x)\right]\,
   \frac{\delta Z_\tau[u]}{\delta u(x)}\,.\eeq
This is a linear, and closed, {\em functional} differential equation. The functional derivative 
${\delta Z_\tau[u]}/{\delta u(x)}$ plays the role of an `annihilation operator' (it reduces
by one the number of factors of $u$), the term quadratic in $u$ describes the `gain'
effect associated with the branching process $x\to (zx, (1-z)x)$, whereas the negative
term linear in $u$ is the corresponding `loss' effect.

By taking one functional derivative w.r.t. $u(x)$ in \eqn{eqZ} and evaluating the result 
at $u(x)=1$, one immediately recovers \eqn{eqD0} for $D(x,\tau)$. By taking two
such derivatives, one finds an equation for $D^{(2)}(x,x',\tau)$ which after simple manipulations
takes the form shown in \eqn{eqD2}.

\section{Computing the gluon pair density}
\label{app:D2}

In this appendix we shall derive Eq.~(\ref{D2exact}) for the gluon pair density.
Our starting point is Eq.~(\ref{D2sol}), that we shall compute for 
the simplified kernel  ${\cal K}_0(z)\equiv 1/{[z(1-z)]^{3/2}}$. Accordingly,
the Green's function $D(x,\tau)$ is given by \eqn{Dexact}, 
while the source term takes the form (recall \eqn{eqD2}) 
\beq
S(x_1,x_2,\tau')=\sqrt{\frac{x_1+x_2}{x_1 x_2}}\,D(x_1+x_2,\tau')=
\frac{\tau'}{\sqrt{x_1 x_2}(1-x_1-x_2)^{3/2}}\ \exp\left\{-\frac{\pi(\tau')^2}{1-x_1-x_2}\right\}\,.
\eeq
Using this together with simple algebraic manipulations, Eq.~(\ref{D2sol}) can be
 rewritten as
\begin{equation}
D^{(2)}(x,x',\tau)=
\int_0^\tau\frac{\,\rmd\tau'(\tau-\tau')^2\tau'}{\sqrt{xx'}}\int_x^1\frac{\,\rmd x_1}{(x_1-x)^{3/2}}\int_{x'}^{1-x_1}
\frac{\,\rmd x_2}{(x_2-x')^{3/2}}\frac{\rme^{-\pi(\tau-\tau')^2\left(\frac{1}{x_1-x}+\frac{1}{x_2-x'}\right)-\frac{\pi(\tau')^2}{1-x_1-x_2}}}{(1-x_1-x_2)^{3/2}}\,.
\label{eq:D2expl}
\end{equation} 
We focus first on the integration in $x_2$ and denote
\begin{equation}
\label{Fdef}
F(1-x_1,x',\tau',\tau)\equiv\int_{x'}^{1-x_1}\frac{\,\rmd x_2}{(x_2-x')^{3/2}(1-x_1-x_2)^{3/2}}\,
\rme^{-\frac{\pi(\tau')^2}{1-x_1-x_2}-\frac{\pi(\tau-\tau')^2}{x_2-x'}}\,.
\end{equation}
After the change of variables $u=\frac{x_2-x'}{1-x_1-x_2}$, we see that the only combination of energy fractions that the integral is sensitive to is $(1-x_1-x')$\,:
\begin{equation}
F(1-x_1,x',\tau',\tau)=\frac{\rme^{-\frac{\pi[(\tau')^2+(\tau-\tau')^2]}{1-x_1-x'}}}{(1-x_1-x')^2}\int_0^\infty\frac{\,\rmd u(u+1)}{u^{3/2}}\,\rme^{-\frac{\pi(\tau')^2u}{1-x_1-x'}-\frac{\pi(\tau-\tau')^2}{(1-x_1-x')u}}\,.
\end{equation}
The integration in $u$ can be done analytically, by using
\begin{equation}
I_1(a,b)\equiv \int_0^\infty\frac{\,\rmd u}{\sqrt{u}}\,\rme^{-\frac{a}{u}-bu}=\sqrt{\frac{\pi}{b}}\,\rme^{-2\sqrt{ab}}\,.
\label{eq:I1}
\end{equation}
Differentiating the above with respect to $a$, we obtain another useful relation,
\begin{equation}
-\frac{\partial}{\partial a}I_1(a,b)=\int_0^\infty\frac{\,\rmd u}{u^{3/2}}\,\rme^{-\frac{a}{u}-bu}=\sqrt{\frac{\pi}{a}}\,\rme^{-2\sqrt{ab}} \,.
\end{equation}
By combining the two above equations, we deduce 
\begin{equation}
\int_0^\infty\frac{\,\rmd u(u+1)}{u^{3/2}}\rme^{-\frac{\pi(\tau')^2u}{1-x_1-x'}-\frac{\pi(\tau-\tau')^2}{(1-x_1-x')u}}=\frac{\tau\sqrt{1-x_1-x'}}{\tau'(\tau-\tau')}\,\rme^{-\frac{2\pi\tau'(\tau-\tau')}{1-x_1-x'}}\,,
\end{equation}
and therefore
\begin{equation}
F(1-x_1,x',\tau',\tau)=\frac{\tau}{\tau'(\tau-\tau')(1-x_1-x')^{3/2}}\rme^{-\frac{\pi\tau^2}{1-x_1-x'}}\,.
\end{equation}
We are left with the integration in $x_1$, which now reads
\begin{equation}
D^{(2)}(x,x',t)=\frac{\tau}{\sqrt{xx'}}\int_0^\tau\,\rmd\tau'(\tau-\tau')\int_x^{1-x'}\frac{\,\rmd x_1}{(x_1-x)^{3/2}(1-x_1-x')^{3/2}}\,\rme^{-\frac{\pi(\tau-\tau')^2}{x_1-x}-\frac{\pi\tau^2}{1-x_1-x'}}\,.
\end{equation}
This integral over $x_1$ is very similar to the one  over $x_2$ that we have just  performed, cf. \eqn{Fdef}:
\begin{equation}
D^{(2)}(x,x',t)=\frac{\tau}{\sqrt{xx'}}\int_0^\tau\,\rmd\tau'(\tau-\tau')F(1-x',x,\tau,2\tau-\tau')=\int_0^\tau\,\rmd\tau'(2\tau-\tau')\frac{\rme^{-\frac{\pi(2\tau-\tau')^2}{1-x-x'}}}{\sqrt{xx'}(1-x-x')^{3/2}}\,.
\end{equation}
The last integral in $\tau'$ is easy to do by using $\int\rmd x \,x\,\rme^{-x^2}=-\frac{\rme^{-x^2}}{2}$ and yields
 the result in Eq.~(\ref{D2exact}).

\section{Computing energy fluctuations}
\label{app:X2}

In this appendix we present in some detail the calculations leading to Eqs.~(\ref{X2A}) and (\ref{X2B}). 
For convenience, we introduce the notations
\beq
 \langle X^2(\tau)\rangle_A\equiv \int_0^1\,\rmd x\int_0^1\,\rmd x'D^{(2)}(x,x',\tau)\,,\qquad
 \langle X^2(\tau)\rangle_B\equiv \int_0^1\,\rmd xxD(x,\tau)\,,
 \eeq 
 and we start by computing $\langle X^2(\tau)\rangle_A$ using Eq.~(\ref{D2exact}) for $D^{(2)}(x,x',\tau)$\,:
\begin{equation}
\langle X^2(\tau)\rangle_A=\frac{1}{2\pi}\int_0^1\frac{\,\rmd x}{\sqrt{x}}\int_0^{1-x}\frac{\,\rmd x'}{\sqrt{x'(1-x-x')}}\left(\rme^{-\frac{\pi\tau^2}{1-x-x'}}-\rme^{-\frac{4\pi\tau^2}{1-x-x'}}\right)\,.
\label{eq:XAfirst}
\end{equation}
The integration in $x'$ can be simplified by making the change of variable $u=\frac{x'}{1-x-x'}$\,:
\begin{equation}
\langle X^2(\tau)\rangle_A=\frac{1}{2\pi}\int_0^1\frac{\,\rmd x}{\sqrt{x}}\left[I\left(\frac{\pi\tau^2}{1-x}\right)-I\left(\frac{4\pi\tau^2}{1-x}\right)\right]\,,
\end{equation}
where
\begin{equation}
I(x)=\rme^{-x}\int_0^\infty\frac{\,\rmd u}{\sqrt{u}(u+1)}\rme^{-ux}\,.
\end{equation}
Using the result in Eq.~(\ref{eq:I1}) we see that $\frac{\rmd I}{\rmd x}=-\sqrt{\frac{\pi}{x}}\rme^{-x}$ which implies
\begin{equation}
I(x)=-\int_x^\infty\,\rmd x'\frac{\rmd I}{\rmd x'}=\sqrt{\pi}\int_0^\infty\frac{\,\rmd x'}{\sqrt{x'}}\rme^{-x'}=2\sqrt{\pi}\int_{\sqrt{x}}^\infty\,\rmd t\rme^{-t^2}=\pi(1-\text{erf}(\sqrt{x}))\,,
\end{equation}
and therefore
\begin{equation}
\langle X^2(\tau)\rangle_A=\frac{1}{2}\int_0^1\frac{\,\rmd x}{\sqrt{x}}\left[\text{erf}\left(\frac{2\sqrt{\pi}\tau}{\sqrt{1-x}}\right)-\text{erf}\left(\frac{\sqrt{\pi}\tau}{\sqrt{1-x}}\right)\right]\,.
\label{eq:XAsecond}
\end{equation}
The remaining integral in $x$ becomes simpler if one first takes a derivative w.r.t. $\tau$\,:
\begin{equation}
\frac{\rmd\langle X^2(\tau)\rangle_A}{\rmd\tau}=\int_0^1\frac{\,\rmd x}{\sqrt{x(1-x)}}\left(2\rme^{-\frac{4\pi\tau^2}{1-x}}-\rme^{-\frac{\pi\tau^2}{1-x}}\right)\,.
\end{equation}
This integral is very similar to the one we have performed over $x'$ and gives
\begin{equation}
\frac{\rmd\langle X^2(\tau)\rangle_A}{\rmd\tau}=\pi(1+\text{erf}(\sqrt{\pi}\tau)-2\text{erf}(2\sqrt{\pi}\tau))\,.
\end{equation}
By also using  $\int\rmd x \,\text{erf}(x)=x\,\text{erf}(x)+\frac{\rme^{-x^2}}{\sqrt{\pi}}$ we get
\begin{equation}
\langle X^2(\tau)\rangle_A=\pi\tau+\sqrt{\pi}\left(\frac{\rme^{-\pi\tau^2}-1}{\sqrt{\pi}}+\sqrt{\pi}\tau 
\,\text{erf}(\sqrt{\pi}\tau)\right)-\sqrt{\pi}\left(\frac{\rme^{-4\pi\tau^2}-1}{\sqrt{\pi}}+2\sqrt{\pi}\tau \,\text{erf}(2\sqrt{\pi}\tau)\right)\,,
\end{equation}
which is equivalent to Eq.~(\ref{X2A}).

Now we move to the computation of $\langle X^2(\tau)\rangle_B$. Using \eqn{Dexact} for $D(x,\tau)$,
this is rewritten as 
\begin{equation}
\label{eq:XB1}
\langle X^2(\tau)\rangle_B=\tau\int_0^1\frac{\,\rmd x \sqrt{x}}{(1-x)^{3/2}}\,\rme^{-\frac{\pi\tau^2}{1-x}}\,.
\end{equation}
Once again, this integral can be simplified by the change of variables $u=\frac{x}{1-x}$\,; one finds
\begin{equation}
\langle X^2(\tau)\rangle_B=\tau G(\pi\tau^2)
\label{eq:XBfirst}
\end{equation}
where
\begin{equation}
G(a)\equiv \int_0^\infty\frac{\,\rmd u\sqrt{u}}{u+1}\rme^{-a(u+1)}\,.
\end{equation}
It is easy to check that $\frac{\rmd G}{\rmd a}=-\frac{\sqrt{\pi}}{2a^{3/2}}\rme^{-a}$, which implies
\begin{equation}
G(a)=\frac{\sqrt{\pi}}{2}\int_a^\infty\frac{\,\rmd a'}{(a')^{3/2}}\rme^{-a'}=\sqrt{\pi}\int_{\sqrt{a}}^\infty\frac{\,\rmd t}{t^2}\rme^{-t^2}\,.
\label{eq:Ga}
\end{equation}
The last integral can be done using integration by parts and the result is
\begin{equation}
G(a)=\sqrt{\frac{\pi}{a}}\,\rme^{-a}-\pi[1-\text{erf}(\sqrt{a})]\,.
\label{eq:Garesult}
\end{equation}
Combining this with Eq.~(\ref{eq:XBfirst}) gives the result in Eq.~(\ref{X2B}).

In order to get more physical intuition about the physical meaning of these results,
it is interesting to see what are the integration regions which contribute up to $\order{\tau^4}$
in the small time regime at $\tau\ll 1$, or $\pi \tau^2\ll 1$.  Our purpose is to substantiate the
physical discussion in Sect.~\ref{sec:dis}.

It is easy to check that
the leading and subleading contributions to $\langle X^2(\tau)\rangle_A$, of $\order{\tau}$
and respectively $\order{\tau^2}$, come from integrating over $x$ and $x'$ with
 $1-x-x'\sim\pi\tau^2$. Indeed, values such that $1-x-x'\ll\pi\tau^2$
 are strongly suppressed by the exponentials in Eq.~(\ref{eq:XAfirst}).  In the region
where $1-x-x'\gg \pi\tau^2$, we can expand the exponentials in Eq.~(\ref{eq:XAfirst}) 
and  then we are led to integrations of the type
\begin{equation}
\int_0^1\frac{\,\rmd x}{\sqrt{x}}\int_0^{1-x}\frac{\,\rmd x'}{\sqrt{x'}(1-x-x')^{1/2+n}}\,,
\end{equation}
with $n\ge 1$. (Notice that the term with $n=0$ has cancelled between the two exponentials.)
But clearly, these integrals have an endpoint singularity at $1-x-x'=0$ (for any $n\ge 1$),
which in the original integral in  Eq.~(\ref{eq:XAfirst}) was cut off by the exponentials at
values $x$ and $x'$ such that $1-x-x'\sim\pi\tau^2$. Then the dominant contribution
comes precisely from the would-be singular endpoint, that is, from the pairs $(x,x')$ with
$1-x-x'\sim\pi\tau^2$.

To more finely disentangle contributions
of various orders in $\tau$, let us now have a look at the integral representation in 
Eq.~(\ref{eq:XAsecond}). The leading-order contribution, linear in $\tau$, can be obtained
by expanding the error functions inside the integrand to linear order in their arguments;
this yields $\langle X^2(\tau)\rangle_A\simeq \pi \tau^2$ and also shows that
the relevant values of $x$ obey $x\sim 1/2$. Together with $1-x-x'\sim\pi\tau^2$, this implies
that the term of $\order{\tau}$ is generated by $x\sim x'\sim 1/2$, as anticipated in
Sect.~\ref{sec:dis}. Beyond linear order, the integrand in Eq.~(\ref{eq:XAsecond}) cannot be
expanded anymore, as this would generate non-integrable singularities at $x=1$. This
shows that the contribution of $\order{\tau^2}$ comes from $1-x\sim\pi\tau^2$, which 
together with $1-x-x'\sim\pi\tau^2$ implies that this quadratic piece comes from pairs $(x,x')$
made with the LP plus one soft particle: $x\simeq 1$ and $x'\sim\tau^2$, or vice-versa.
It is furthermore easy to check that the respective contributions to $\langle X^2(\tau)\rangle_B$
have a similar origin: the contribution to \eqn{eq:XB1}
 of $\order{\tau}$ comes from $x\sim 1/2$, whereas that  of $\order{\tau^2}$ comes from
 $x\sim\tau^2$.

It seems difficult to precisely characterize the integration region in $x$ and $x'$ which
contributes to $\langle X^2(\tau)\rangle_A$ at $\order{\tau^4}$~: this is a subsubleading
piece, hence already the regions previously identified can give (subleading) contributions
of $\order{\tau^4}$. It is however clear that, to this order, one opens a new physical channel,
which refers to the production of pairs $(x,x')$ where both particles are relatively soft,
so that they are produced with a probability of order one, yet at the same time they are 
hard enough to contribute to the energy-weighted pair density: these conditions select
$x\sim x'\sim \tau^2$. To see this, let us focus on the contribution of the soft modes  to 
Eq.~(\ref{eq:XAfirst}). This can be estimated by introducing an upper cutoff  
$x_\Lambda\sim \tau^2$ in the integrals over $x$ and $x'$, while at the same
time keeping only the first non-trivial term, of $\order{\tau^2}$, in the expansion of the exponentials
(this expansion makes sense in the presence of the cutoff, since $1-x-x'\simeq 1\gg\tau^2$). One thus
finds, to parametric accuracy,
\begin{equation}
\langle X^2(\tau)\rangle_A|_{x<x_\Lambda}\sim \frac{3\tau^2}{2}\int_0^{x_\Lambda}\frac{\rmd x}{\sqrt{x}}\int_0^{x_\Lambda}\frac{\rmd x'}{\sqrt{x'}}+\mathcal{O}(\tau^2x_\Lambda^2)=6\tau^2x_\Lambda+\mathcal{O}(\tau^2x_\Lambda^2)\,,
\end{equation}
which is indeed of $\order{\tau^4}$ for the physically motivated choice $x_\Lambda\sim \tau^2$.

\section{Computing multiplicity fluctuations}
\label{app:N2}

In this appendix, we shall compute the second factorial moment of the gluon number distribution,
$\langle N(N-1)\rangle(\tau, x_0)$, by explicitly performing the integrations over $x$ and $x'$
which appear in Eq.~(\ref{avNN}). After also using \eqn{D2exact} for $D^{(2)}(x,x',\tau)$,
the starting point integral reads
\begin{equation}\label{app:NN1}
\langle N(N-1)\rangle(\tau, x_0)
=\frac{1}{2\pi}\int_{x_0}^{1-x_0}\frac{\,\rmd x}{x^{3/2}}\int_{x_0}^{1-x}\frac{\,\rmd x'}{(x')^{3/2}\sqrt{1-x-x'}}\left(\rme^{-\frac{\pi\tau^2}{1-x-x'}}-\rme^{-\frac{4\pi\tau^2}{1-x-x'}}\right)\,.
\end{equation}
The integral over $x'$ can be simplified via the change of variables $u=\frac{x'}{1-x-x'}$, which gives
\begin{equation}\label{app:NN2}
\langle N(N-1)\rangle=\frac{1}{2\pi}\int_{x_0}^{1-x_0}\frac{\,\rmd x}{x^{3/2}(1-x)}\int_{u_0}^\infty\frac{\,\rmd u}{u^{3/2}}\left(\rme^{-\frac{\pi\tau^2(u+1)}{1-x}}-\rme^{-\frac{4\pi\tau^2(u+1)}{1-x}}\right)\,,
\end{equation}
where $u_0=\frac{x_0}{1-x-x_0}$. The integral over $u$ is similar to that previously computed
in Eq.~(\ref{eq:Ga}), hence
\begin{equation}
\langle N(N-1)\rangle=\frac{\tau}{\pi}\int_{x_0}^{1-x_0}\frac{\,\rmd x}{x^{3/2}(1-x)^{3/2}}\left[\rme^{-\frac{\pi\tau^2}{1-x}}G\left(\frac{\pi\tau^2u_0}{1-x}\right)-2\rme^{-\frac{4\pi\tau^2}{1-x}}G\left(\frac{4\pi\tau^2u_0}{1-x}\right)\right]\,.
\end{equation}
After substituting the explicit form for the function $G(a)$ from \eqn{eq:Garesult}, this result
can be decomposed into two pieces, corresponding to the 2 terms in the r.h.s. of
Eq.~(\ref{eq:Garesult}). Specifically,
$\langle N(N-1)\rangle=\langle N(N-1)\rangle_A+\langle N(N-1)\rangle_B$, where
\begin{align}\label{app:NNA}
\langle N(N-1)\rangle_A=&\frac{1}{\pi\sqrt{x_0}}\int_{x_0}^{1-x_0}\frac{\,\rmd x}{x^{3/2}}
\frac{\sqrt{1-x-x_0}}{1-x}\left(\rme^{-\frac{\pi\tau^2}{1-x-x_0}}-\rme^{-\frac{4\pi\tau^2}{1-x-x_0}}\right)
\end{align}
and respectively
\begin{align}\label{app:NNB}
\langle N(N-1)\rangle_B=&
\tau\int_{x_0}^{1-x_0}\frac{\,\rmd x}{x^{3/2}(1-x)^{3/2}}
\Bigg\{2\rme^{-\frac{4\pi\tau^2}{1-x}}\left[1-\text{erf}\left(\frac{2\tau\sqrt{\pi x_0}}{\sqrt{(1-x)(1-x-x_0)}}\right)\right]\nonumber\\
&\qquad\qquad\qquad\qquad\quad
 -\rme^{-\frac{\pi\tau^2}{1-x}}\left[1-\text{erf}\left(\frac{\tau\sqrt{\pi x_0}}{\sqrt{(1-x)(1-x-x_0)}}\right)\right]\Bigg\}\,.
\end{align}
In order to understand better the physical origin of the different terms it is useful to note that
\begin{equation}\label{app:G2}
G(a)=\sqrt{\pi}\rme^{-a}\int_{\sqrt{a}}^\infty\frac{\,\rmd t}{t^2}+\sqrt{\pi}\int_{\sqrt{a}}^\infty\frac{\,\rmd t}{t^2}(\rme^{-t^2}-\rme^{-a})\,,
\end{equation}
where the new integration variable $t$ is related to the previous variables $u$ and $x'$ appearing in 
\eqref{app:NN2} and respectively \eqref{app:NN1} (say, for the first exponential within the integrand there), via
\beq
t^2 =\frac{\pi\tau^2}{1-x}\,u= \frac{\pi\tau^2}{1-x}\,\frac{x'}{1-x-x'}\,.\eeq
(For the second exponential within the integrand one should simply replace $\tau\to 2\tau$.)
The first integral in \eqn{app:G2} corresponds to the first term in Eq.~(\ref{eq:Garesult})
 and is dominated by $t\sim\sqrt{a}$; this implies that the integral over $x'$ that led to the result
in \eqn{app:NNA} is controlled by its lower limit $x'\sim x_0$. Similarly, 
the second integral in  \eqn{app:G2} corresponds to the second term in Eq.~(\ref{eq:Garesult}) and is controlled 
by $t\sim 1$; accordingly, the integral over $x'$ leading to \eqn{app:NNB}  is controlled by
 $\frac{x'}{1-x-x'}\sim \frac{1-x}{\pi\tau^2}$. 

We have not been able to analytically compute the remaining integrals over $x$, but we managed
to obtain good approximations in two interesting limits, to be described in what follows.

\subsection{The case $x_0\ll \pi\tau^2\ll 1$}

We start by computing $\langle N(N-1)\rangle_A$ according to \eqn{app:NNA}. 
The region at $1-x-x_0\ll \pi\tau^2$ being exponentially suppressed, it is enough to consider
the domain $1-x-x_0\gtrsim \pi\tau^2$, meaning $1-x\gtrsim \pi\tau^2+x_0\simeq \pi\tau^2$,
where we have used the fact that $\pi\tau^2\gg x_0$. In other terms, we can substitute
$1-x-x_0\simeq 1-x$ up to higher order terms, so the integration becomes
\begin{equation}
\langle N(N-1)\rangle_A\,\simeq\,
\frac{1}{\pi\sqrt{x_0}}\int_{x_0}^{1}\frac{\,\rmd x}{x^{3/2}\sqrt{1-x}}\left(\rme^{-\frac{\pi\tau^2}{1-x}}-\rme^{-\frac{4\pi\tau^2}{1-x}}\right)\,.
\end{equation}
After the change of variables $u=\frac{x}{1-x}$, the integral is similar to the one in Eq.~(\ref{eq:Ga}), so we have

\begin{align}\label{app:NNA1}
\langle N(N-1)\rangle_A\,\simeq\,
&\frac{1}{\pi\sqrt{x_0}}\Bigg[\rme^{-\pi\tau^2}\left(\frac{2}{\sqrt{x_0}}\rme^{-\pi\tau^2x_0}-2\pi\tau\big[1-\text{erf}(\sqrt{\pi x_0}\tau)\big]\right)\nn
&\qquad -\rme^{-4\pi\tau^2}\left(\frac{2}{\sqrt{x_0}}\rme^{-4\pi\tau^2x_0}-4\pi\tau\big[1-\text{erf}(\sqrt{4\pi x_0}\tau)\big]\right)\Bigg]\nonumber\\
&= \frac{6\tau^2}{x_0}+\frac{2\tau}{\sqrt{x_0}}+\mathcal{O}\left(\frac{\tau^4}{x_0}\right)\,.
\end{align}
The term quadratic in ${\tau}/{\sqrt{x_0}}$, which is the dominant term in this regime, comes
from the lower limit $x\sim x_0$. The subleading term, linear in ${\tau}/{\sqrt{x_0}}$, arises by
integrating over $\frac{x}{1-x}\sim\frac{1}{\pi\tau^2}$, which for $\tau\ll 1$ is tantamount to $x\sim 1-\pi\tau^2$.
Recalling that the whole contribution to $\langle N(N-1)\rangle_A$ comes from $x'\sim x_0$, we conclude
that the dominant term ${6\tau^2}/{x_0}$ is generated by the region where both measured gluons
live close to the infrared cutoff, $x\sim x'\sim x_0$, whereas the subleading term
${2\tau}/{\sqrt{x_0}}$ comes from the pairs made with the LP plus a soft gluon: $x\simeq 1-\pi\tau^2$
and $x'\sim x_0$, or vice-versa.

Consider similarly the other contribution,  $\langle N(N-1)\rangle_B$ in \eqn{app:NNB}. 
The terms involving error functions give only subleading contributions: indeed, the arguments
of the error functions are much smaller than one except in the region where the integrand is already exponentially suppressed. The pole of the integrand at $x=1$ is cut off by the exponential at $1-x\sim\pi\tau^2$, hence
its contribution to the final result is subleading too. We conclude that the dominant contribution comes
from the lower limit $x=x_0$, due to the $1/x^{3/2}$ pole. So, we can write
\begin{equation}\label{app:NNB1}
\tau\int_{x_0}^{1-x_0}\frac{\,\rmd x}{x^{3/2}(1-x)^{3/2}}\left(2\rme^{-\frac{4\pi\tau^2}{1-x}}-\rme^{-\frac{\pi\tau^2}{1-x}}\right)= \tau\int_{x_0}^\infty\frac{\,\rmd x}{x^{3/2}}+\mathcal{O}(1)= \frac{2\tau}{\sqrt{x_0}}+\mathcal{O}(1)\,.
\end{equation}
We recall that this piece $\langle N(N-1)\rangle_B$ has been obtained by integrating over $x'$ with
$\frac{x'}{1-x-x'}\sim\frac{1-x}{\pi\tau^2}$. Together with $x\simeq x_0$, this implies $x'\simeq 1-x_0-\pi\tau^2\simeq 1-\pi\tau^2$. 
Hence, the interpretation of the linear term in \eqn{app:NNB1} is the same as for the respective
term in  \eqn{app:NNA1} : this is the effect of the pairs made with the LP plus a soft gluon.

Our final result in this regime reads
\begin{equation}\label{ap:NNsmall}
\langle N(N-1)\rangle=\frac{6\tau^2}{x_0}+\frac{4\tau}{\sqrt{x_0}}+\mathcal{O}\left(1,\frac{\tau^4}{x_0}\right)\,,
\end{equation}
where the term quadratic in $\tau$ is generated by processes like those in Fig. \ref{fig:correl} (where $x\sim x'\sim x_0$), while the linear term is coming from diagrams like that in Fig. \ref{fig:var} right
(with $x\sim 1-\pi\tau^2$ and $x'\sim x_0$).

\subsection{The case $\pi\tau^2\ll x_0\ll 1$}

In this regime, it is convenient to rewrite the integral in  \eqn{app:NNA} as 
\begin{align}\label{app:NNAlarge}
\langle N(N-1)\rangle_A\,=\,
&\frac{3\tau^2}{\sqrt{x_0}}\int_{x_0}^{1-x_0}\frac{\,\rmd x}{x^{3/2}(1-x)\sqrt{1-x-x_0}}\nonumber\\
&
+\frac{1}{\pi\sqrt{x_0}}\int_{x_0}^{1-x_0}\frac{\,\rmd x}{x^{3/2}}\frac{\sqrt{1-x-x_0}}{1-x}
\left(\rme^{-\frac{\pi\tau^2}{1-x-x_0}}-\rme^{-\frac{4\pi\tau^2}{1-x-x_0}}-\frac{3\pi\tau^2}{1-x-x_0}\right)\,.
\end{align}
This is convenient since 
the expression in the second line can be neglected to the accuracy of interest: indeed,
one can check that the respective contribution is of order ${\tau^3}/{x_0^{3/2}}$
and hence it is parametrically smaller than the terms that we shall keep in the final result
(see \eqn{ap:NNlarge} below). The integral in the first of \eqn{app:NNAlarge} 
line can be simplified with the change of variables $u=\frac{x}{1-x-x_0}$ :
\begin{equation}\label{app:NNAlarge1}
\langle N(N-1)\rangle_A\simeq \,
\frac{3\tau^2}{\sqrt{x_0}}\int_{u_0}^\infty\frac{\,\rmd u(u+1)}{u^{3/2}(1+ux_0)}=\frac{3\tau^2}{\sqrt{x_0}}\int_{u_0}^\infty\frac{\,\rmd u}{u^{3/2}}+\frac{3\tau^2(1-x_0)}{\sqrt{x_0}}\int_{u_0}^\infty\frac{\,\rmd u}{\sqrt{u}}\frac{1}{1+ux_0}\,,
\end{equation}
with $u_0=\frac{x_0}{1-2x_0}$. The first integral in the r.h.s. is dominated by $u\sim u_0$,
i.e. $x\sim x_0$, and gives
\begin{equation}
\frac{3\tau^2}{\sqrt{x_0}}\int_{u_0}^\infty\frac{\,\rmd u}{u^{3/2}}=\frac{6\tau^2}{x_0}+\mathcal{O}(\tau^2\sqrt{x_0})\,.
\end{equation} 
Recalling that $x'\sim x_0$ for $\langle N(N-1)\rangle_A$, it is clear that the above contribution 
comes from gluon pairs where both particles are soft.
In this regime where $x_0$ is relatively large, perturbation theory is guaranteed to work,
so the above result, which is quadratic in $\tau$, 
can be interpreted in terms of processes involving exactly 2 branchings
and leading to two final gluons with $x\sim x'\sim x_0$. Clearly, these are the 3 processes
illustrated in Fig.~\ref{fig:2g}.

The second integral in \eqn{app:NNAlarge1} 
 can be computed via the change of variables $t^2=ux_0$ :
\begin{equation}
\frac{3\tau^2(1-x_0)}{\sqrt{x_0}}\int_{u_0}^\infty\frac{\,\rmd u}{\sqrt{u}}\frac{1}{1+ux_0}=\frac{6\tau^2(1-x_0)}{x_0}\int_{\sqrt{u_0x_0}}^\infty\frac{\,\rmd t}{t^2+1}=\frac{3\pi\tau^2}{x_0}+\mathcal{O}(\tau^2)\,.
\label{eq:pi}
\end{equation}
This is controlled by $t\sim 1$, which implies $u\sim 1/x_0$, or $x\sim 1-2x_0$. This contribution too
is generated by processes involving exactly 2 branchings, that is, the 3 processes shown  
in Fig.~\ref{fig:2g}. But in this case, one counts the pairs made with the LP (with $x\sim 1-2x_0$)
and one of the 2 soft emitted gluons (with $x'\sim x_0$), while the other one is not measured. However, this contribution does not survive in the final result, as we shall shortly see.

Consider now the second contribution $\langle N(N-1)\rangle_B$, cf. \eqn{app:NNB}.
The upper limit $1-x_0 > x$ on the integration variable in \eqn{app:NNB}
implies $1-x > x_0 \gg\pi\tau^2$, hence the arguments of the exponentials within
the integrand are bound to be small and can be replaced by 1 to the accuracy of
interest (up to corrections of order $\tau^3$). We divide the calculation into two parts,
with the first part defined by those terms in the integrand which do not involve
the error functions. Hence, the first part involves the following integral (once again, we set
$u=\frac{x}{1-x-x_0}$)
\begin{equation}
\tau\int_{x_0}^{1-x_0}\frac{\,\rmd x}{x^{3/2}(1-x)^{3/2}}\simeq
\tau\int_{u_0}^\infty\frac{\,\rmd u (u+1)}{u^{3/2}(1+x_0u)^{3/2}}\,,
\end{equation}
where $u_0=\frac{x_0}{1-2x_0}$ and we have used $1-x_0\simeq 1$. 
We can now manipulate the last integral to obtain
\begin{equation}\label{app:int}
\tau\int_{u_0}^\infty\frac{\,\rmd u}{u^{3/2}}+\tau\int_{u_0}^\infty\frac{\,\rmd u}{\sqrt{u}(1+x_0u)^{3/2}}+\tau\int_{u_0}^\infty\frac{\,\rmd u}{u^{3/2}}\left(\frac{1}{\sqrt{1+x_0u}}-1\right)\,.
\end{equation}
The first integral above is dominated by $x\sim x_0$ and can be easily computed:
\begin{equation}
\tau\int_{u_0}^\infty\frac{\,\rmd u}{u^{3/2}}=\frac{2\tau}{\sqrt{x_0}}+\mathcal{O}(\tau\sqrt{x_0})\,.
\end{equation}
Remembering that the result in \eqn{app:NNB} comes 
from the region $\frac{x'}{1-x-x'}\sim\frac{1-x}{\pi\tau^2}$, we see that $x'\simeq 1-x_0-\pi\tau^2\simeq
1-x_0$. This result is linear in $\tau$, meaning that it describes a single emission by the LP, of a gluon
with energy fraction $x\sim x_0$. Now we look at the second integral in \eqn{app:int}:
\begin{equation}
\tau\int_{u_0}^\infty\frac{\,\rmd u}{\sqrt{u}(1+x_0u)^{3/2}}\simeq \frac{2\tau}{\sqrt{x_0}}\int_{x_0}^\infty\frac{\,\rmd t}{(t^2+1)^{3/2}}=\frac{2\tau}{\sqrt{x_0}}+\mathcal{O}(\tau\sqrt{x_0})\,,
\label{app:real}
\end{equation}
(we set $t^2=x_0u$),
where the term linear in $\tau$ comes from $t\sim 1$, or $x\sim 1-2x_0$. This furthermore
implies (after also using $\frac{x'}{1-x-x'}\sim\frac{1-x}{\pi\tau^2}$) that $x'\sim 2x_0$. Once again,
this contribution comes from a single emission by the LP, of a gluon
with energy $x'\sim 2x_0$. Finally the third integral in \eqn{app:int}
can be seen to give a subleading contribution of order $\tau\sqrt{x_0}$.


We now turn to the last remaining piece of the calculation, that involving that part
of the integrand in \eqn{app:NNB} which is proportional to the error function. We recall
that the arguments of the exponentials there should be replaced by 1.
It can be checked that up to terms of order $\tau^3$ it is correct the replace each of the 
error functions by the first term in their small-argument expansions. This gives
\begin{align}
-6\tau^2\sqrt{x_0}\int_{x_0}^{1-x_0}\frac{\,\rmd x}{x^{3/2}(1-x)^2\sqrt{1-x-x_0}}=-\frac{6\tau^2\sqrt{x_0}}{1-x_0}\int_{u_0}^\infty\frac{\,\rmd u(u+1)^2}{u^{3/2}(1+ux_0)^2}\,,
\end{align}
where in the second equality we made the change of variables $u=\frac{x}{1-x-x_0}$, so $u_0=\frac{x_0}{1-2x_0}$. We can divide the last integral into two pieces,
\begin{equation}
-\frac{6\tau^2\sqrt{x_0}}{1-x_0}\int_{u_0}^\infty\frac{\,\rmd u\sqrt{u}}{(1+ux_0)^2}-\frac{6\tau^2\sqrt{x_0}}{1-x_0}\int_{u_0}^\infty\frac{\,\rmd u((u+1)^2-u^2)}{u^{3/2}(1+ux_0)^2}\,.
\label{eq:2pieces}
\end{equation}
The first of these integrals can be done analytically with the change of variables $ux_0=t^2$ :
\begin{equation}\label{app:virt}
-\frac{12\tau^2}{x_0(1-x_0)}\int_{\sqrt{u_0x_0}}^\infty\frac{\,\rmd t t^2}{(t^2+1)^2}=-\frac{3\pi\tau^2}{x_0}+\mathcal{O}(\tau^2)\,.
\end{equation}
The above integral is dominated by $t\sim 1$, which implies $x\sim 1-2x_0$ and therefore
$x'\sim 2x_0$. 
This particular contribution is negative, hence it should be interpreted as a `virtual' (or `loss') correction
to the one-gluon emission in \eqn{app:real}, that is, a loop correction expressing the reduction
in the probability for having the LP. Moreover, this negative piece has the right magnitude
to cancel the contribution computed in Eq.~(\ref{eq:pi}), which we recall refers to processes with
2 gluons emissions out of which only one is measured. Such a `real' vs. `virtual' cancellation was
in fact to be expected, it occurs whenever some of the gluons produced in the final state are
not measured. The second integral in Eq.~(\ref{eq:2pieces}) gives a contribution 
of order $\tau^2$, which is negligible to the present accuracy.


Putting everything together we finally deduce
\begin{equation}\label{ap:NNlarge}
\langle N(N-1)\rangle=\frac{4\tau}{\sqrt{x_0}}+\frac{6\tau^2}{x_0}+\mathcal{O}\left(\frac{\tau^3}{x_0^{3/2}}\right)\,,
\end{equation}
which is formally similar with the previous result in eq. (\ref{ap:NNsmall}), except that now the
term linear in $\tau$ dominates over the quadratic one (and the whole result is much smaller than 1).


\providecommand{\href}[2]{#2}\begingroup\raggedright\endgroup

\end{document}